\documentclass[prd,aps,preprint,nofootinbib,]{revtex4-1}

\usepackage{epsfig}
\usepackage{amsmath,amssymb}
\usepackage{graphicx,subfigure}
\usepackage{color}

\newcommand{\be}{\begin{equation}}
\newcommand{\ee}{\end{equation}}
\newcommand{\bea}{\begin{eqnarray}}
\newcommand{\eea}{\end{eqnarray}}
\def\lla{\left\langle}
\def\rra{\right\rangle}
\def\za{\alpha}
\def\zb{\beta}
\def\ssc{\scriptscriptstyle}
\def\lsim{\mathrel{\raise.3ex\hbox{$<$\kern-.75em\lower1ex\hbox{$\sim$}}} }
\def\gsim{\mathrel{\raise.3ex\hbox{$>$\kern-.75em\lower1ex\hbox{$\sim$}}} }

\begin{document}

\preprint{{\vbox{\hbox{NCU-HEP-k042}
\hbox{Jul 2011}
\hbox{rev. Nov 2011}
}}}

\vspace*{.5in}
\title{Dynamical Symmetry Breaking with Four-Superfield Interactions
\\[.5in]}

\author{ Gaber Faisel }
\email{gfaisel@cc.ncu.edu.tw}

\affiliation{
Department of Physics and \\
Center for Mathematics and Theoretical Physics,
National Central University, Chung-Li, Taiwan 32054.
}

\author{ Dong-Won Jung }
\email{dwjung@phys.nthu.edu.tw}

\affiliation{
Department of Physics, National Tsing Hua University and \\
Physics Division, National Center for Theoretical Sciences,
Hsinchu, Taiwan, 300
}

\author{ Otto C. W. Kong }
\email{otto@phy.ncu.edu.tw}

\affiliation{Department of Physics and \\
Center for Mathematics and Theoretical Physics,
National Central University, Chung-Li, Tawian, 32054.
}

\maketitle

\vspace*{.5in}
\begin{center}
{\Large Abstract}
\end{center}
\vspace*{.5in}
We investigate the dynamical mass generation resulted from interaction terms with
four chiral superfields. The kind of interactions maybe considered a supersymmetric
generalization of the four-fermion interactions of the classic Nambu--Jona-Lasinio
model. A four-superfield interaction that contains a four-fermion interaction as one
of its component terms has been the standard supersymmetrization of the NJL model for
decades. Recently, we introduced a holomorphic variant with a dimension five
interaction term instead. The latter is a main target of the present analysis.
With the introduction of a new perspective on the superfield gap equation, we
derive it for each one of the four-superfield interactions, using the supergraph
technique. Through analyzing solutions to the gap equations, we illustrate the
dynamical generation of superfield
Dirac mass, including a supersymmetry breaking part. A dynamical symmetry breaking
generally goes along with the dynamical mass generation, for which a bi-superfield
condensate is responsible. The explicit illustration of dynamical symmetry breaking
from the holomorphic dimension five interaction is reported for the first time.
It has rich and novel features, which would be easily missed without the superfield
approach developed here. We also discuss
the nature of the  bi-superfield condensate and its role of the effective Higgs
superfield picture for both cases, illustrating their difference.
Note that such a holomorphic quark superfield interaction term can successful account
for the electroweak symmetry breaking with Higgs superfields as composites.


\newpage

\section{Introduction}
Dynamical mass generation and symmetry breaking is a very interesting theoretical topic
with important phenomenological applications. In the early studies of mechanism
for spontaneous symmetry breaking, Nambu adopted the idea of Cooper pairing \cite{Nobel}
to construct a classic model of dynamical mass generation and symmetry breaking. This
is the Nambu--Jona-Lasinio (NJL) model \cite{Nambu:1961tp}, with a strong attractive
four-fermi interaction. It can be shown through the analysis of the nonperturbative
gap equation that the interaction induces a bi-fermion vacuum condensate which serves
as the source for the fermion Dirac mass. The condensate naturally breaks whatever
symmetry the bi-fermion configuration carries a nontrivial quantum number of. The
classic example is the (Dirac fermion) chiral symmetry, which was Nambu's first
concern \cite{Nobel,N60}. After the Standard Model was generally established, the
exact mechanism of electroweak symmetry breaking became a problem of paramount
importance in the phenomenological domain. It is still open till today. It was pointed
out by Nambu \cite{Nambu} that for a sufficiently heavy top quark, an NJL model of top
condensate can give rise to electroweak symmetry breaking. The top quark, however,
turns out to be not heavy enough \cite{tSM,Chung:2005yz}.

Supersymmetry is another important theme in modern physics. The idea of constructing
a supersymmetric version of the NJL model was introduced in 1982 \cite{BL}. A dimension
six four-superfield interaction containing the NJL four-fermion interaction was the
basic feature. The superfield gap equation analysis however showed no nontrivial mass
solution. The model supplemented with soft supersymmetric breaking mass terms was later
established, through an effective potential analysis of its effective theory with auxiliary
Higgs superfield introduced, to success in giving dynamical mass generation \cite{BE}.
A direct gap equation analysis from the original model Lagrangian, however, has not
been available. On the side of phenomenological applications, a gauged version of the
kind of model was presented for electroweak symmetry breaking, giving the supersymmetric
Standard Model as the low-energy effective theory \cite{CLB,92}. Supersymmetry
fixes the unnatural fine tuning of the four-fermion coupling required in the
NJL model of top condensation. Phenomenological viability of the model has been severely
unfavorable cornered with the relatively small top mass value determined and constraint
on the $\tan\!\beta$ parameter \cite{034}.

In a re-examination of the supersymmetrization of the NJL model, it was realized that
there is a natural alternative to the one given in Refs.\cite{BL,BE}, as elaborated in
Ref.\cite{035}. The alternative was presented in Ref.\cite{034}, together with an explicit
model version that can give rise to electroweak symmetry breaking. The new model has a
dimension five four-superfield interaction instead. The interaction is a superpotential
term, hence holomorphic. It was named the holomorphic supersymmetric
Nambu--Jona-Lasinio (HSNJL) model, while the the name supersymmetric Nambu--Jona-Lasinio
(SNJL) model is kept for the dimension six version. The HSNJL model for electroweak
symmetry breaking is a gauged version with two composite superfields. It also gives rise to
the supersymmetric Standard Model as the low-energy effective theory. A first step
renormalization group analysis was presented to illustrate the basic compatibility of
the current experimental constraints on the latter model with the HSNJL model features.
We are interested in establishing the dynamical mass generation for the generic
HSNJL model structure, with a direct gap equation analysis. More generally, we seek
to obtain gap equation results for both the dimension five and dimension six
four-superfield interactions. We success in doing that, only with the development of a
new perspective on the superfield gap equation. We obtain gap equations and dynamical
mass generation results for both cases with generic couplings. For the case with
the dimension five interaction, our analysis here mostly focused on the simplest
version of a HSNJL model, with one superfield composite. Generalization to the
two composite case should be straight forward. For the dimension six case, we obtain
a supersymmetric breaking part of the dynamical mass generation not available before.

Our calculational framework is based on the supergraph techniques of
Grisaru, Siegel and Ro\v{c}ek \cite{GSR}, with extension to fully accommodate
supersymmetric breaking effect as developed by Miller \cite{M}, Helayel-Neto \cite{H},
Scholl \cite{S}, and others. In a way, we take
it a step further. All couplings, including mass terms, in the superfield
Lagrangian are considered like constant superfields, {\it i.e.} with in general
both a supersymmetric parts and a (soft) supersymmetry breaking parts. The effective
action may then be considered as a superfield functional with explicit
dependence on the Grassmannian coordinates $\theta$ and $\bar{\theta}$, hence
also having supersymmetric and supersymmetry breaking parts. In particular,
we will calculate the proper self-energy with its supersymmetric part and
supersymmetry breaking part to obtain the gap equation for a model as a pair
of coupled equations involving both the corresponding supersymmetric and
supersymmetry breaking parts of the Dirac mass parameter. Note that the
supersymmetric part contribute to the usual fermion Dirac mass as well as
mass-squared for the two scalar field components, while the supersymmetry
breaking part contributes to the mass mixing between the scalars. One will
see that the approach is powerful for a comprehensive analysis of the general
case of the parameter value. If one considers only the parameter $m$ it will
leads to the wrong, or at least in complete, answer.

Our gap equation analysis will be presented in Sec.II, which is the main content
of the paper. In Sec.III, we discussed some details of the mass generation and
symmetry breaking features, after which we conclude in Sec.IV.

\section{Dynamical Mass Generation --- Gap Equation Analysis\label{gapsec}}
We start with the following Lagrangian density for what one wants to have as a Dirac
pair of chiral (`quark') superfields,
$\Phi_\pm (y,\theta)=A_\pm(y) + \sqrt{2} {\theta} \psi_\pm(y) + {\theta}^2 F_\pm(y)$ :
\bea
{\cal L}  =
\int\!\!  d^4 \theta ~
 \bigg[ \left( \Phi_+^\dagger \Phi_+
  + \Phi_-^\dagger \Phi_- \right) (1-\Delta)
+ \big({\mathcal M} \, \Phi_+ \Phi_- \,\delta^2\!(\bar{\theta})
+ h.c. \big) \bigg]
\;\;+ \;\;{\cal L}_I \;.
\eea
Here, $\Delta = \tilde{m}^2 \theta^2 {\bar \theta}^2$ characterizes a soft
supersymmetry breaking mass-squared $\tilde{m}^2$ for the scalar fields
$A_\pm$ and ${\mathcal M}$ a superfield Dirac mass parameter. An important point
to note is that ${\mathcal M}$ should be considered like a constant superfield
with a supersymmetric as well as a supersymmetry breaking part \cite{M,S}. We write
\be
{\mathcal M} = m - {\theta}^2 \eta \;,
\ee
where $m$ is the usual (supersymmetric) Dirac mass and $\eta$ its supersymmetry breaking
counterpart. In general, $m$ and $\eta$ should be taken as complex. At least a relative
phase between the two cannot be rotated away. The former corresponds to Dirac mass for the
fermion pair $\psi_\pm$ and $|m|^2$ contributions to both $A_\pm$ mass-squared, while
the supersymmetry breaking part $\eta$ gives (so-called left-right) mass mixing between
the $A_\pm$ pair, and does not correspond to a mass eigenvalue. The Dirac mass term is a
superpotential term; the $\int \! d^4 \theta \,\delta^2\!(\bar{\theta})$
integral reduces to an $\int \! d^2 \theta$ integral in the commonly written form.
The crucial step in our analysis is to write the effective action as
$\Gamma \equiv \Gamma(\Phi_+,\Phi_-,\Phi^{\dagger}_+,\Phi^{\dagger}_-,\theta,\bar{\theta})$
accordingly, where the explicit dependence on $\theta$ and $\bar{\theta}$ allows
supersymmetry breaking parts to be included. We are interested in the quadratic part
$\Gamma^{(2)}_{+-}(p,\theta)$ in $\Gamma$, with
\be
 \Gamma  = \int  \frac{d^4 p}{2\pi^4} \int \!\! d^2 \! \theta \,
  \Phi_+(-p,\theta) \, \Gamma^{(2)}_{+-}(p,\theta^2)  \,
 \Phi_-(p,\theta)  +h.c.
+ \cdots \;,
\label{effact}
\ee
where
\be
\Phi_{\pm}(p,\theta)  =  \int d^4 \! x ~ e^{-ip.x}
  \Phi_{\pm}(x,\theta) \;.
\ee
The $\Gamma^{(2)}_{+-}(p,\theta^2)$ function again contains in general a
scalar part and a part with  ${\theta}^2$, in exact analog to the parameter
${\mathcal M}$. The former is supersymmetric while the latter is supersymmetry
breaking, corresponding to the $\eta$ term in ${\mathcal M}$
\footnote{
The effective action is commonly written as
\[
 \Gamma  = \int  \frac{d^4 p}{2\pi^4} \int \!\! d^4 \theta \,
  \Phi_+(-p,\theta,\bar{\theta}) \, \Gamma^{(2)}_{+-}(p,\theta,\bar{\theta})  \,
 \Phi_-(p,\theta,\bar{\theta})  \, \delta^2\!(\bar{\theta}) +h.c.
+ \cdots \;,
 \]
even for a unbroken supersymmetric theory (same in Ref.\cite{BL}).
However,  $\Gamma ^{(2)}_{+-}$  has in that case no real dependence on $\theta$ and
$\bar{\theta}$ \cite{Gates:1983nr}. The term in the effective action is of course chiral,
hence the way we write it here, adding the explicit $\theta^2$  dependence.
}.

As a self-consistent Hartree approximation for the dynamically generated nonzero
${\mathcal M}$, the interaction Lagrangian density ${\cal L}_I$ is taken to contain a
$ -\big[{\mathcal M} \, \Phi_+ \Phi_- \delta^2\!(\bar{\theta}) + h.c. \big]$
term and at least an extra true interaction term which is to be the true origin
of the nonzero Dirac mass parameter. One looks for nontrivial solution for
${\mathcal M}$ from
the equation
\be
\left. \Gamma ^{(2)}_{+-}(p,\theta^2)\right|_{\mbox{\tiny on-shell}}  = 0
\ee
which is equivalent to the vanishing of the proper self-energy
\be
\left.\Sigma_{+-}(p,\theta^2)\right|_{\mbox{\tiny on-shell}}  =0
\ee
from diagrams produced by ${\cal L}_I$. With the ${\mathcal M}$ term in ${\cal L}_I$,
we have
\be \label{gap}
- {\mathcal M} =  \left. \Sigma^{({\mbox \tiny loop})}_{+-}(p,\theta^2)
 \right|_{\mbox{\tiny on-shell}} \;,
\ee
where $\Sigma^{({\mbox \tiny loop})}_{+-}$ denotes the lowest order contributions
to the proper self-energy from loop diagrams involving the true interaction, {\it i.e.}
either one of the four-superfield interactions we will consider. As for ${\mathcal M}$
and $\Gamma^{(2)}_{+-}$, $\Sigma^{({\mbox \tiny loop})}_{+-}$ includes plausiby a
supersymmetry breaking part. The above is the gap equation. However, we have exactly
extended the usual gap equation as an equation for the Dirac mass $m$ to one for
${\mathcal M}$, which may then be interpreted as two coupled equations for $m$ and $\eta$.

With the framework for the calculation outlined, now we come to the proper self-energy
diagrams through supergraph analysis. First of all, we need the superfield propagators in
the generic framework. We obtained, within the formulation of
Grisaru, Siegel and Ro\v{c}ek \cite{GSR},
\bea
\langle T(\Phi_{\pm}(1) \Phi_{\pm}^\dagger(2)) \rangle
&=&
\frac{-i}{p^2+|m|^2} \delta^4_{\ssc 12}
-\frac{i}{[(p^2+|m|^2+\tilde{m}^2)^2-|\eta|^2]}
\left( \eta \bar{m}  \theta_{\!\ssc 1}^2
+ \bar{\eta} m\bar{\theta_{\!\ssc 1}}^2\right) \,\delta^4_{\ssc 12}
\nonumber \\ &&
+\frac{i\; [\tilde{m}^2(p^2+|m|^2+\tilde{m}^2)
-|\eta|^2]}{(p^2+|m|^2)[(p^2+|m|^2+\tilde{m}^2)^2-|\eta|^2]}
  \left[ |m|^2 \theta_{\!\ssc 1}^2 \bar{\theta_{\!\ssc 1}}^2
+ \frac{D_{\!\ssc 1}^2 \theta_{\!\ssc 1}^2 \bar{\theta_{\!\ssc 1}}^2
\overline{D}_{\!\ssc 1}^2}{16} \right] \,\delta^4_{\ssc 12} \;,
\nonumber \\
&& \\
\lla T(\Phi_+(1) \Phi_-(2)) \rra &=&
\frac{i \, \bar{m}}{p^2(p^2+|m|^2)} \frac{D_{\!\ssc 1}^2}{4} \delta^4_{\ssc 12}
\nonumber \\&&
-\frac{i}{[(p^2+|m|^2+\tilde{m}^2)^2-|\eta|^2]} \!
\left[ \frac{\bar{\eta} \, D_{\!\ssc 1}^2  \bar{\theta_{\!\ssc 1}}^2}{4}
- \frac{\eta |m|^2 \, D_{\!\ssc 1}^2 {\theta_{\!\ssc 1}}^2}{4p^2} \!\right]
\!\delta^4_{\ssc 12}
\nonumber  \\&&
+ \frac{i\, \bar{m} \; [\tilde{m}^2(p^2+|m|^2
+\tilde{m}^2)-|\eta|^2]}{(p^2+|m|^2)[(p^2+|m|^2+\tilde{m}^2)^2-|\eta|^2]} \;
\left[\frac{D_{\!\ssc 1}^2 \theta_{\!\ssc 1}^2 \bar{\theta_{\!\ssc 1}}^2}{4}
+  \frac{\bar{\theta_{\!\ssc 1}}^2 \theta_{\!\ssc 1}^2 D_{\!\ssc 1}^2}{4}
\!\right] \delta^4_{\ssc 12} \;.
\label{sprop} \eea
where $\delta^4_{\ssc 12}=\delta^4(\theta_{\!\ssc 1}-\theta_{\!\ssc 2})$.
Our expressions here agree with Refs.\cite{S,H}.

Next, we introduce the interactions of interest that are expected to lead to
nontrivial $\Sigma^{({\mbox \tiny loop})}_{+-}$. Consider the dimension six
four-superfield interaction
\be \label{d6}
g^2 \int\!\! d^4 \theta \, \Phi_+^\dagger\Phi_-^\dagger \Phi_+\Phi_- \,
(1- \tilde{m}_{\!\ssc C}^2 \theta^2 {\bar \theta}^2 )
\ee
coming with a supersymmetry breaking part. This interaction gives the SNJL model,
here extended to include the supersymmetry breaking $\tilde{m}_{\ssc C}^2$ part.
The dimension five four-superfield interaction is given by
\be \label{d5}
-\frac{G}{2} \int\!\! d^4 \theta \, \Phi_+\Phi_-\Phi_+\Phi_- \,
(1+ B \theta^2)\, \delta^2\!(\bar{\theta}) \;.
\ee
It is really a superpotential term, as indicated by the $\delta^2\!(\bar{\theta})$,
hence holomorphic. This HSNJL model is proposed as an alternative supersymmetrization
of the NJL model. The two four-superfield interactions are the focus of our interest.

With the above, we are ready to implement the supergraph evaluation of
$\Sigma^{({\mbox \tiny loop})}_{+-}(p, \theta^2)$ at one-loop level. We use a
technique from Miller \cite{M} on one-loop tadpole, extending it to the proper
self-energy diagram. The technique can be considered as re-writing the
effective action as
\be
 \Gamma  = \int  \frac{d^4 p}{2\pi^4} \left. \int \!\! d^2 \!\theta \,
  \Phi_+(-p,\theta_1)  \left[ \int \!\! d^2 \!\bar{\theta} \,
\Gamma^{(2)}_{+-}(p,\theta^2) \, \delta^2\!(\bar{\theta}) \right]
 \Phi_-(p,\theta_2)   \right|_{\theta_1=\theta_2=\theta}  +h.c.
+ \cdots \;,
\label{Meffact}
\ee
splitting the vertex with $\theta_1$ and $\theta_2$ distinct from $\theta$ in
the evaluation of the diagrams before finally enforcing the equal limit. The
proper self-energy diagram is to be taken as an integrand over $d^4 \theta$,
with amplitude at the ${\theta_1=\theta_2=\theta}$ limit contributing to
$\Sigma^{({\mbox \tiny 1loop})}_{+-}(p, \theta^2)$ as given schematically
by
\be
\left. \int \!\! d^2 \!\bar{\theta} \,
\big[\mbox{Amplitude}\big] \right|_{\theta_1=\theta_2=\theta}
\longrightarrow
\int \!\! d^2 \!\bar{\theta} \,
\Sigma^{({\mbox \tiny 1loop})}_{+-}(p,\theta^2) \, \delta^2\!(\bar{\theta}) \;.
\ee
Note that at the one-loop level, $\Sigma^{({\mbox \tiny 1loop})}_{+-}$ is
actually independent of the external momentum $p$, with a loop momentum integral
to be evaluated with a cut-off $\Lambda$. The on-shell condition for the
gap equation [{\it c.f.} Eq.(\ref{gap})] is trivial.

\subsection{Results from the dimension six interaction}
For dimension six interaction, the $g^2$ vertex gives at one-loop level the proper
self-energy diagram of Fig.~1. As clear from the diagram, a
$\Phi_+^\dagger \Phi_-^\dagger $ propagator is involved in the corresponding proper
self-energy $\Sigma^{({\mbox \tiny fig1})}$, which is independent of the external
momentum $p$.  The propagator as from (the conjugate of) Eq.(\ref{sprop}) has three
terms, the last two each has two parts (from the two fractions inside the big bracket).
We present our results on the contribution from each part separately schematically as
\[
\Sigma^{({\mbox \tiny fig1})}
= \Sigma^{({\mbox \tiny fig1})}_1 + \Sigma^{({\mbox \tiny fig1})}_{2a}
 + \Sigma^{({\mbox \tiny fig1})}_{2b}
 + \Sigma^{({\mbox \tiny fig1})}_{3a}
 + \Sigma^{({\mbox \tiny fig1})}_{3b} \;,
\]
in order to illustrate the roles of the supersymmetric and supersymmetry breaking
parts of the coupling, the  mass parameter ${\mathcal M}$, and the propagator on the
$\Sigma^{({\mbox \tiny fig1})}$. Reader may then read off directly from the results
the contributions to the the supersymmetric ($m$) and supersymmetry breaking ($\eta$)
parts of gap equation,
\[
- {\mathcal M} =  \Sigma^{({\mbox \tiny fig1})}
\]
in this case, as to be presented below.

Through the term by term supergraph calculations of the partial amplitudes, we obtain
\bea
\int \!\! d^2 \!\bar{\theta} \,
\Sigma^{({\mbox \tiny fig1})}_{1} \, \delta^2\!(\bar{\theta})
&=&
\int \!\! d^2 \!\bar{\theta} \,0  \;,
\nonumber \\
\int \!\! d^2 \!\bar{\theta} \,
\Sigma^{({\mbox \tiny fig1})}_{2a} \, \delta^2\!(\bar{\theta})
&=&
\int \!\! d^2 \!\bar{\theta} \,(-\eta g^2)
(1- \tilde{m}_{\!\ssc C}^2 \theta^2 {\bar \theta}^2 ) \,
I_2(|m|^2,\tilde{m}^2, |\eta|, \Lambda^2)  \;,
\nonumber \\
\int \!\! d^2 \!\bar{\theta} \,
\Sigma^{({\mbox \tiny fig1})}_{2b} \, \delta^2\!(\bar{\theta})
&=&
\int \!\! d^2 \!\bar{\theta} \,0  \;,
\nonumber \\
\int \!\! d^2 \!\bar{\theta} \,
\Sigma^{({\mbox \tiny fig1})}_{3a} \, \delta^2\!(\bar{\theta})
&=&
\int \!\! d^2 \!\bar{\theta} \, m g^2
 (1- \tilde{m}_{\!\ssc C}^2 \theta^2 {\bar \theta}^2 ) \,
I_1(|m|^2,\tilde{m}^2, |\eta|, \Lambda^2) \, \bar{\theta}^2
\;,
\nonumber \\
\int \!\! d^2 \!\bar{\theta} \,
\Sigma^{({\mbox \tiny fig1})}_{3b} \, \delta^2\!(\bar{\theta})
&=&
\int \!\! d^2 \!\bar{\theta} \,m g^2
 (1- \tilde{m}_{\!\ssc C}^2 \theta^2 {\bar \theta}^2 ) \,
I_1(|m|^2,\tilde{m}^2, |\eta|, \Lambda^2) \, \bar{\theta}^2
  \;,
\eea
where
\bea
I_1(|m|^2, \tilde{m}^2, |\eta|, \Lambda^2)
&=&  \int\frac{d^4k}{(2\pi)^4} \frac{ [\tilde{m}^2(k^2+|m|^2
+\tilde{m}^2)-|\eta|^2]}{(k^2+|m|^2)[(k^2+|m|^2+\tilde{m}^2)^2-|\eta|^2]}
\nonumber \\
= &&\!\! \!\!\!\!  \!\!\!\!
\frac{1}{16\pi^2} \left[ \frac{1}{2}(|m|^2+\tilde{m}^2)
\ln{\frac{(|m|^2+\tilde{m}^2+\Lambda^2)^2-|\eta|^2}{(|m|^2+\tilde{m}^2)^2-|\eta|^2}}
-|m|^2 \ln{\frac{(|m|^2+\Lambda^2)}{|m|^2}} \right. \nonumber \\
 && + \left. |\eta|
\left(\tanh^{-1}\frac{|m|^2+\tilde{m}^2+\Lambda^2}{|\eta|}
-\tanh^{-1}\frac{|m|^2+\tilde{m}^2}{|\eta|}
\right) \right] \;,
\nonumber \\
I_2(|m|^2, \tilde{m}^2, |\eta|, \Lambda^2)
&=&  \int\frac{d^4k}{(2\pi)^4}\frac{1}{(k^2+|m|^2+\tilde{m}^2)^2-|\eta|^2}
\nonumber \\
= &&\!\! \!\!\!\!  \!\!\!\!
\frac{1}{16\pi^2} \left[ \frac{1}{2}
\ln{\frac{(|m|^2+\tilde{m}^2+\Lambda^2)^2-|\eta|^2}{(|m|^2+\tilde{m}^2)^2-|\eta|^2}}
\right. \nonumber \\
&& \!\!\!\! + \left. \frac{|m|^2+\tilde{m}^2}{|\eta|}
\left(\tanh^{-1}\frac{|m|^2+\tilde{m}^2+\Lambda^2}{|\eta|}
-\tanh^{-1}\frac{|m|^2+\tilde{m}^2}{|\eta|} \right) \right] .
\label{I12} \eea

One can see that the two loop function have a
built-in constraint that $(|m|^2+\tilde{m}^2)^2-|\eta|^2>0$. This
is nothing other than the condition for the left-right mixing mass
between the scalars to be small enough to avoid giving a {tachyonic}
mass eigenvalue. For the gap equation (\ref{gap}) with
$\Sigma^{({\mbox \tiny fig1})}$, we hence obtain
\bea
 m &=& 2 m  g^2  I_1(|m|^2, \tilde{m}^2, |\eta|, \Lambda^2) \;,
\nonumber \\
\eta &=& - \eta \, g^2 \tilde{m}_{\!\ssc C}^2 \, {I}_2(|m|^2, \tilde{m}^2, |\eta|, \Lambda^2) \;.
\label{gap6}
\eea
Note that the equations for $m$ and $\eta$ are somewhat decoupled, which we will
see is not the case with the dimension five interaction.
\begin{figure}[!t]
\begin{center}
\includegraphics[height=4.0cm,width=10.0cm]{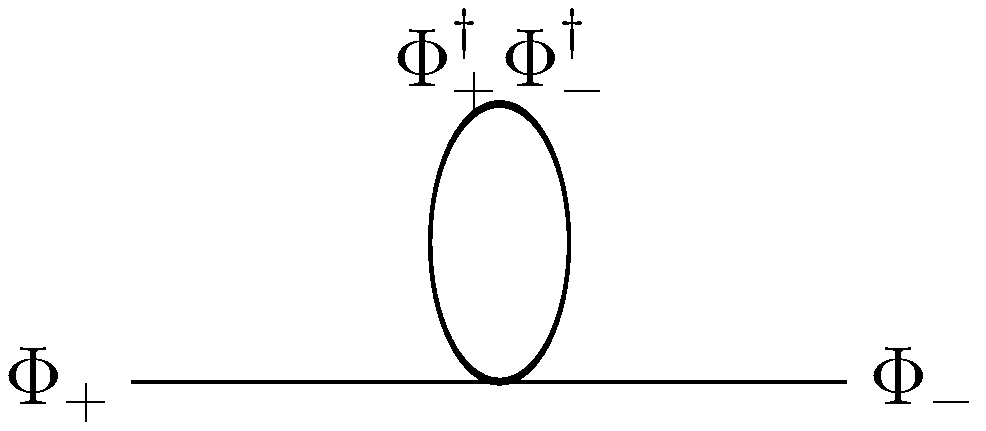}
\end{center}
\caption{\small Superfield diagram for proper self-energy $\Sigma_{+-}(p, \theta^2)$
with the dimension six four-superfield interaction.}
\vspace*{.2in}
\hrule
\end{figure}

A couple of remarks are in order. Firstly, if the parameter $\tilde{m}_{\!\ssc C}^2$
vanishes, we have $\eta =0$ and hence the gap equation for $m$ reduces to
\be
m = 2 m \, g^2 I(|m|^2,\tilde{m}^2,\Lambda^2) \;,
\ee
where
\be
 I(|m|^2,\tilde{m}^2,\Lambda^2)
=\frac{1}{16\pi^2} \left[ |m|^2 \ln{
\frac{|m|^2(|m|^2+\tilde{m}^2+\Lambda^2)}
{(|m|^2+\tilde{m}^2)(|m|^2+\Lambda^2)} } +\tilde{m}^2\ln \left(
1+\frac{\Lambda^2}{|m|^2+\tilde{m}^2} \right) \right] \;.
\ee
The equation gives a supersymmetric Dirac mass $m$ which vanishes for $\tilde{m}^2 =0$.
Note that the case with both $\tilde{m}_{\!\ssc C}^2$ and $\tilde{m}^2$ being zero
corresponds to the SNJL model with an exactly supersymmetric Lagrangian \cite{BL}.
On the other hand, taking the limit  $\tilde{m}\rightarrow \infty$ where the scalar particles
of $\Phi_{\pm}$ become heavy and decoupled,  $m$ becomes the simple Dirac fermion/quark mass
which then satisfies the equation
\be
m= \frac{m g^2}{8\pi^2} \big[ \Lambda^2 + |m|^2 \ln{|m|^2} -|m|^2 \ln(\Lambda^2 + |m|^2) \big] \;,
\label{gapeqq}
\ee
giving
\be
(g^2)^{-1}= \frac{\Lambda^2}{8\pi^2}
\left[ 1 - \frac{|m|^2}{\Lambda^2} \, \ln{\frac{\Lambda^2 }{|m|^2}}
+ O(1/\Lambda^4) \right] \;.
\ee
This is the standard NJL result for the pure fermionic model usually given with
and extra $N_c$ factor for the $\psi_{\pm}$ fermions as colored quarks.
For the SNJL case with nonzero $\tilde{m}^2$, the gap equation for $m$ has been
obtained in  Ref.\cite{BE} through a component field effective potential analysis of the
low energy effective Lagrangian after the introduction of two auxiliary (Higgs/composite)
superfields. Their result is
\be
\frac{2 m g }{\Lambda^2}  \left[ \frac{8\pi^2}{ g^2\Lambda^2}
-\frac{|m|^2+\tilde{m}^2}{\Lambda^2}
\ln\left( 1+\frac{\Lambda^2}{|m|^2+\tilde{m}^2} \right)
+\frac{|m|^2}{\Lambda^2} \ln \left( 1+{\frac{\Lambda^2}{|m|^2}}  \right)
 \right]=0  \;.
\ee
One can easily see, with a little algebra, that it agrees exactly with our result. Note that
nontrivial solution for $m$ exists for the coupling constant satisfying the inequality \cite{BE}
\be
 g^2   > \frac{8\pi^2}{\tilde{m}^2
\ln \left( 1+\frac{\Lambda^2}{\tilde{m}^2} \right)}   \;,
\ee
generating a mass for the Dirac fermion pair. As we will also illustrate in the next
section, the mass term comes from a bi-superfield condensate, which contains a bi-fermion
part, that breaks the chiral symmetry in the original Lagrangian.

The case without supersymmetry breaking corresponds to
$\tilde{m}^2 =\tilde{m}_{\!\ssc C}^2=0$.
In that case, a supergraph analysis has been performed going to two-loop
evaluation of $\Sigma^{({\mbox \tiny loop})}_{+-}(p,\theta)$ \cite{BL}.
No nontrivial solution for $m$ exists. While we do not have serious doubt on
that result, we note that their superfield gap equation analysis
did not include the equation for $\eta$, the supersymmetric breaking part, as we do.
Our result here, however, explicitly establishes that, up to the one-loop level,
vanishing $\tilde{m}_{\!\ssc C}^2$ necessarily implies zero $\eta$ no matter what
value $\tilde{m}^2$ has. Formally speaking, such fully general gap equation result
still have to be obtained for the two-loop analysis; and nontrivial solution for $\eta$
will imply spontaneous supersymmetry breaking.

To further illustrate the power of our general gap equation results, we report
also result for a scenario on the other extreme where $m=0$ but $\eta \ne 0$
solution for Eq.(\ref{gap6}). Naively, one enforces zero $m$ in the the $I_2$
integral of the equation for $\eta$. Nontrivial solution for the latter
exists under the condition
\bea
\frac{1}{16\pi^2} \left[\ln{\left(1+\frac{\Lambda^2}{ \tilde{m}^2}\right)}
-\frac{\Lambda^2}{\Lambda^2+\tilde{m}^2}\right]
\leq \frac{1}{-g^2\tilde{m}_{\!\ssc C}^2} <
\frac{1}{16\pi^2}\ln{\left(1+\frac{\Lambda^2}{2\tilde{m}^2}\right)} \;,
\eea
details of the analysis behind which we leave to appendix~A.
The last part of the inequality comes from an analysis similar to that
of the condition for nonotrivial $m$ under $\eta=0$. The magnitude of the
responsible coupling,  ${g^2\tilde{m}_{\!\ssc C}^2}$ here, has to be big enough.
The other part of the inequality is actually from
$|\eta| \leq (m^2+\tilde{m}^2)$ beyond which there will be a tachyonic
scalar mass eigenvalue. Note that one always needs a negative
${\tilde{m}_{\!\ssc C}^2}$ for nontrivial $\eta$ solution. The parameter
is to be appreciated as the soft mass term for the composite superfield,
as we will show explicitly in the next section.

Next we turn to the dimension five holomorphic interaction case, which
actually shows an amazingly strong interplay between $m$ and $\eta$.

\subsection{Results from the dimension five interaction}
For the dimension five interaction, the G vertex gives at one-loop level a
diagram only slightly different from the previous case, as Fig.~2. The propagator
$\Phi_+ \Phi_-$ is involved instead. Again, we write schematically
\[
\Sigma^{({\mbox \tiny fig2})}
= \Sigma^{({\mbox \tiny fig2})}_1 + \Sigma^{({\mbox \tiny fig2})}_{2a}
 + \Sigma^{({\mbox \tiny fig2})}_{2b}
 + \Sigma^{({\mbox \tiny fig2})}_{3a}
 + \Sigma^{({\mbox \tiny fig2})}_{3b} \;.
\]
Our calculations give the partial amplitude from the various parts
of the propagator in Eq.(\ref{sprop}) as
\bea \int \!\! d^2
\!\bar{\theta} \, \Sigma^{({\mbox \tiny{fig2}})}_{1} \, \delta^2\!(\bar{\theta})
&=& \int\!\! d^2 \!\bar{\theta} \,0 \, \delta^2\!(\bar{\theta}) \;,
\nonumber \\
\int \!\! d^2 \!\bar{\theta} \, \Sigma^{({\mbox \tiny{fig2}})}_{2a} \,
\delta^2\!(\bar{\theta})
&=& \int\!\! d^2 \!\bar{\theta} \,  \, \frac{\bar{\eta} G}{2} (1+ B
\theta^2) \,
 I_2(|m|^2,\tilde{m}^2, |\eta|, \Lambda^2)\,
\delta^2\!(\bar{\theta}) \;,
\nonumber \\
\int \!\! d^2 \!\bar{\theta} \, \Sigma^{({\mbox \tiny{fig2}})}_{2b} \,
\delta^2\!(\bar{\theta})
&=& \int\!\! d^2 \!\bar{\theta} \,0 \, \delta^2\!(\bar{\theta}) \;,
\nonumber \\
\int \!\! d^2 \!\bar{\theta} \, \Sigma^{({\mbox \tiny{fig2}})}_{3a} \,
\delta^2\!(\bar{\theta})
 &=& \int\!\! d^2 \!\bar{\theta} \, \frac{-\bar{m} G}{2} (1+ B \theta^2) \,
I_1(|m|^2,\tilde{m}^2, |\eta|, \Lambda^2) \, {\theta}^2 \,
\delta^2\!(\bar{\theta}) \;,
\nonumber \\
\int \!\! d^2 \!\bar{\theta} \, \Sigma^{({\mbox \tiny{fig2}})}_{3b} \,
\delta^2\!(\bar{\theta})
&=& \int\!\! d^2 \!\bar{\theta} \,
 \frac{-\bar{m} G}{2} (1+ B \theta^2) \,
I_1(|m|^2,\tilde{m}^2, |\eta|, \Lambda^2) \, {\theta}^2 \,
\delta^2\!(\bar{\theta}) \;.
\eea
where $I_1(|m|^2, \tilde{m}^2, |\eta|, \Lambda^2) $ and
$I_2(|m|^2, \tilde{m}^2, |\eta|, \Lambda^2)$ are the same loop
integrals as in Eq.(\ref{I12}). That gives the gap equation with
$\Sigma^{({\mbox \tiny {fig2}})}$ as
\bea m &=&
\frac{\bar{\eta} G}{2}\; I_2(|m|^2,\tilde{m}^2,|\eta|, \Lambda^2)\;,
\nonumber \\
\eta &=&  \bar{m} G \;  I_1(|m|^2,\tilde{m}^2,|\eta|, \Lambda^2)
- \frac{\bar{\eta} G B}{2} \; I_2(|m|^2,\tilde{m}^2,|\eta|, \Lambda^2) \;.
\eea
\begin{figure}[!t]
\begin{center}
\includegraphics[height=4.0cm,width=10.0cm]{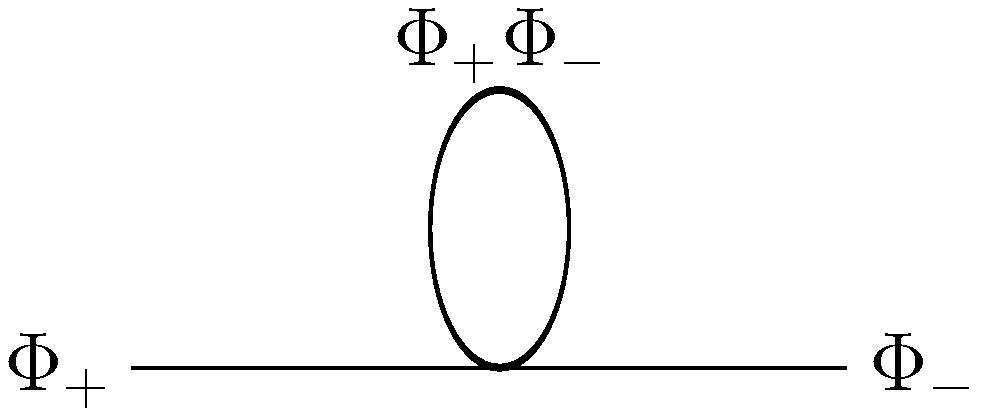}
\end{center}
\caption{\small Superfield diagram for proper self-energy $\Sigma_{+-}(p, \theta^2)$
with the dimension five four-superfield interaction.}
\vspace*{.2in}
\hrule
\end{figure}

The first thing to note in the gap equation result is the
important fact that the equations for $m$ and $\eta$ are completely coupled.
If one naively drop $\eta$ from consideration, for instance, one will not
see any nontrivial expression and completely miss the possible dynamical
mass generation. The two parameters will either both have
nontrivial solutions or both vanishing.

Considering only the case of real values for $m$ and $\eta$ under the assumption of a real
and small $B$ value, we find that nontrivial solution exists for large enough G
(taken as real and positive here by convention) satisfying
\be
G > \sqrt{G_0^2 + b^2} + b \sim G_0  + b\;,
\ee
where
\bea
G_0^2
= \frac{512\pi^2}{\tilde{m}^2\ln{\left(1+\frac{\Lambda^2}{\tilde{m}^2}\right)}
              \left[\ln{\left(1+\frac{\Lambda^2}{\tilde{m}^2}\right)}
 -\frac{\Lambda^2}{\tilde{m}^2+\Lambda^2}\right]} \;
\eea
gives the critical $G^2$ for $B=0$, and
\be
b = B \; \frac{8\pi^2}{\tilde{m}^2\ln{\left(1+\frac{\Lambda^2}{\tilde{m}^2}\right)}} \;.
\ee
Details are to be given in appendix~A. Note that $B$ may be positive or negative,
or more generally contains a complex phase. Solution condition for more general cases
is to be further investigated.

\section{Model Features of Mass Generation and Symmetry Breaking}
Let us take a careful look at the mass generation and symmetry breaking structure of the
two simplest models with the different four-superfield interactions. That is best done in each
case with the consideration of the effective theory having the composite superfield as a basic
ingredient. We will see that the HSNJL model has theoretical features at least as
appealing as the NJL model itself, while the SNJL model maybe considered somewhat inferior
in comparison. The model also raises the idea of bi-scalar condensate playing a key role
in (fermion) mass generation and symmetry breaking.

In our HSNJL model with the dimension five four-superfield interaction \cite{034},
the bi-superfield condensate $\lla \Phi_+ \Phi_- \rra$ is considered to be the source
of  mass generation. Hence, whatever symmetry the superfield product carries a nontrivial
quantum number will be spontaneously broken through the dynamics. The symmetry breaking
vacuum condensate can have a supersymmetric part and a supersymmetry breaking part. One
can see easily that the condensate induces mass term ${\mathcal M}$ by explicitly matching
expression (\ref{d5}) to the term as
\[
-\frac{G}{2} \int\!\! d^2 \!\theta \, \lla \Phi_+\Phi_- \rra \Phi_+\Phi_- \,
(1+ B \theta^2 )
\longrightarrow
 \int\!\! d^2 \!\theta \, (m - \eta \, \theta^2) \, \Phi_+\Phi_- \;,
\]
which gives
\bea
m \;(\Phi_+\Phi_-)_F
&=&
-\, G \lla \Phi_+\Phi_- \rra_A \, (\Phi_+\Phi_-)_F  \;,
\nonumber \\
\eta \; (\Phi_+\Phi_-)_A
&=&
G \bigg( \lla \Phi_+\Phi_- \rra_F
+ B \lla \Phi_+\Phi_- \rra_A \bigg)
 (\Phi_+\Phi_-)_A  \;.
\label{m5}
\eea
The supersymmetry breaking nature of $\eta$ is clearly illustrated in the last equation.
It is really a contribution to the so-called left-right mixing of the scalar mass.
It has already been pointed out that the model has the bi-superfield condensate
$\lla \Phi_+\Phi_- \rra_F$ [which contains a bi-fermion part :
$(\Phi_+\Phi_-)_F=A_+F_-+F_+A_--\psi_+\psi_-$] contributing only
to scalar mass (mixing) while it is the bi-scalar condensate $\lla \Phi_+\Phi_- \rra_A$
that gives Dirac fermion mass \cite{035}. Recall that the model actually has no
four-fermion interaction, unlike the case of the SNJL model.

In the SNJL model with the dimension six four-superfield interaction, however,
the story is less straightforward. The interaction term from expression (\ref{d6})
in the presence of the  condensate $\lla \Phi_+ \Phi_- \rra$ reads
\[
g^2 \int\!\! d^4 \theta \, \lla\Phi_+^\dagger\Phi_-^\dagger \rra \Phi_+\Phi_- \,
(1- \tilde{m}_{\!\ssc C}^2 \theta^2 {\bar \theta}^2 ) \;.
\]
The component content is given by
\[
 g^2 \lla\Phi_+^\dagger\Phi_-^\dagger \rra_{\!F} \, (\Phi_+\Phi_-)_F
\qquad\mbox{and} \qquad
 - g^2  \tilde{m}_{\!\ssc C}^2 \,
\lla\Phi_+^\dagger\Phi_-^\dagger \rra_{\!A} \, (\Phi_+\Phi_-)_A  \;.
\]
For the two terms to be recast as components of a superfield Dirac mass term
${\mathcal M}$, we have
\bea
m &=&  g^2 \lla\Phi_+^\dagger\Phi_-^\dagger \rra_{\!F}  \;,
\nonumber \\
\eta &=& g^2  \tilde{m}_{\!\ssc C}^2 \,
\lla\Phi_+^\dagger\Phi_-^\dagger \rra_{\!A} \;.
\label{m6} \eea
The matching of $\lla\Phi_+^\dagger\Phi_-^\dagger \rra$ to ${\mathcal M}$ is hence
in the wrong order, with a switching of the scalar $A$- and auxiliary $F$- parts.
Only through the introduction of an auxiliary Higgs superfield $\Phi_{\!\ssc H}$ other
than the $\Phi_+\Phi_-$ composite one can put the term originated in the K\"ahler potential
as a superpotential term. This twisting of matching the dimension six K\"ahler potential
term to a superpotential term as the Dirac mass is the reason why one cannot take the
$\Phi_+ \Phi_-$ composite to be an effective Higgs superfield directly  \cite{BE}
--- a price to pay to retain a four-fermion interaction.

The very different theoretical structure between the two models will be more transparent
with the effective theory perspective. For the HSNJL model, we have \cite{034}
\bea
{\cal L}_{\mbox{\tiny HSNJL}}^{\mbox{\tiny eff}}
 &=&\int\! d^4\theta  \left( {\Phi}_+^\dagger {\Phi}_+
+ {\Phi}_-^\dagger {\Phi}_- \right) (1-\tilde{m}^2 \theta^2\bar{\theta}^2)
\nonumber \\
&& + \left\{ \int\! d^2\!\theta \left[ \frac{1}{2}  ( \sqrt{\mu_0}{\Phi}_0
+\sqrt{G}  {\Phi}_+ {\Phi}_-) ( \sqrt{\mu_0}{\Phi}_0
+\sqrt{G}  {\Phi}_+ {\Phi}_-)
\right.\right. \nonumber \\ && \left.\left. \qquad
- \frac{G}{2} {\Phi}_+ {\Phi}_- {\Phi}_+ {\Phi}_- \right] (1+B \theta^2)+ h.c. \right\}
\nonumber \\
&=& \int d^4\theta  \left( {\Phi}_+^\dagger {\Phi}_+
+ {\Phi}_-^\dagger {\Phi}_- \right) (1-\tilde{m}^2 \theta^2\bar{\theta}^2)
\nonumber \\ &&
+\left\{\int\! d^2\!\theta \left[ \frac{\mu_0}{2} {\Phi}_0^2
+ \sqrt{\mu_0 G}{\Phi}_0 {\Phi}_+ {\Phi}_- \right] (1+B \theta^2) + h.c.\right\}\;,
\label{L5e}
\eea 
where ${\Phi}_0$ is the auxiliary Higgs superfield that comes out as the composite
\be
{\Phi}_0 = -\sqrt{G/\mu_0}\,{\Phi}_+ {\Phi}_-\;,
\ee
from its own equation of motion. Hence, the Lagrangian differs from the original one only in
having an extra term that vanished by the above equation. The path integral over the auxiliary
superfield is a trivial Gaussian. When it is integrated out in the effective theory, one can
retrieve the generating functional of the original theory up to a constant factor.
This feature mimics exactly that of the NJL model\cite{KK}. As the ${\Phi}_0$
develops a vacuum expection value (VEV), we have the mass
\be
{\mathcal M} = m - \eta \theta^2 = \sqrt{\mu_{0} G} \lla {\Phi}_0 \rra (1+B \theta^2)
\ee
or
\bea
m &=& \sqrt{\mu_{0} G} \lla {\Phi}_0 \rra_A \;,
\nonumber \\
\eta &=&
- \sqrt{\mu_{0} G}  \left(\, \lla {\Phi}_0 \rra_F
+ B \lla {\Phi}_0 \rra_A  \right) \;.
\eea
The result matches directly to that of Eq.(\ref{m5}), again mimicking exactly the
basic feature in the NJL model. Despite having a four-superfield interaction that does
not contain a four-fermion interaction, the HSNJL model with its effective Lagrangian
formulation looks like an exact supersymmetrization of the NJL model.

Situation for the case of the SNJL model is quite a bit more complicated. In the effective
theory picture, the Higgs superfield cannot be obtained as the composite. We have instead
\bea
{\cal L}_{\mbox{\tiny SNJL}}^{\mbox{\tiny eff}}
&=& \int d^4\theta  \left[
 \left( {\Phi}_+^\dagger {\Phi}_+
+ {\Phi}_-^\dagger {\Phi}_- \right) (1-\tilde{m}^2 \theta^2\bar{\theta}^2)
+ {\Phi}_{\!\ssc C}^\dagger {\Phi}_{\!\ssc C}  \,
(1- \tilde{m}_{\!\ssc C}^2 \theta^2 {\bar \theta}^2 ) \right]
\nonumber \\ &&
+\left\{\int\! d^2 \!\theta \; \mu \; {\Phi}_{\!\ssc H}
  \left(  {\Phi}_{\!\ssc C} + g \Phi_+\Phi_-  \right) (1+A \theta^2)
 + h.c.\right\}\;.
\label{L6e}
\eea 
where ${\Phi}_{\!\ssc H}$ is the auxiliary Higgs superfield whose equation of motion
gives the other superfield introduced ${\Phi}_{\!\ssc C}$ as the composite
\be
{\Phi}_{\!\ssc C} = - {g} \Phi_+\Phi_- \;.
\ee
In addition, we have put in an extra supersymmetry breaking part characterized by
parameter $A$ in the superpotential. Unlike the $B$ parameter in the previous model, the
admissible $A$ parameter is arbitrary; it is not even related to any parameter in the
original Lagrangian. Apart from adding to the original Lagrangian the superpotential
term which is constrained to be zero, we actually have to replace the
$g^2 {\Phi}_+^\dagger {\Phi}_-^\dagger {\Phi}_+{\Phi}_-$ in the K\"ahler potential by
${\Phi}_{\!\ssc C}^\dagger {\Phi}_{\!\ssc C}$ too. Moreover, from the point of view of
the original Lagrangian, there is no clear picture on how the ${\Phi}_{\!\ssc H}$, or
rather  $\mu {\Phi}_{\!\ssc H}$, arises out of the dynamics for ${\Phi}_\pm$. It is not a
simple composite. Equations of motion for components of ${\Phi}_{\!\ssc C}$ from the
effective Lagrangian give, for instance,
\bea
\mu F_{\ssc H} +A \,\mu  A_{\ssc H}  &=& - \partial^m\partial_m A_{\ssc C}^*
 - \tilde{m}_{\!\ssc C}^2 \, A_{\ssc C}^* \;,
\nonumber \\
\mu \psi_{\ssc H} &=& -i \sigma^m \partial_m \psi_{\ssc C}^\dagger\;,
\nonumber \\
\mu A_{\ssc H} &=& -F_{\ssc C}^* \;,
\label{c2h}
\eea
respectively. Nevertheless, the Lagrangian in the presence of nontrivial VEV
for ${\Phi}_{\!\ssc H}$ gives
\be
{\mathcal M} = m - \eta \, \theta^2 = {\mu g} \lla {\Phi}_{\!\ssc H} \rra (1+A \theta^2) \;.
\ee
The result needs to be mapped to that of Eq.(\ref{m6}) as
\bea
m &=& {\mu g} \lla {\Phi}_{\!\ssc H} \rra_A
\quad \longrightarrow \quad
- {g} \lla {\Phi}_{\!\ssc C}^\dagger \rra_F \;,
\nonumber \\
\eta &=&
-{\mu g}  \left(\, \lla {\Phi}_{\!\ssc H} \rra_F
+ A \lla {\Phi}_{\!\ssc H} \rra_A  \right)
\quad \longrightarrow \quad
- {g}\, \tilde{m}_{\!\ssc C}^2 \lla {\Phi}_{\!\ssc C}^\dagger \rra_A \;.
\eea
The matching is consistent with Eq.(\ref{c2h}). Contrary to a claim in Ref.\cite{92},
we see that nonzero $\eta$ and hence nonzero $\lla {\Phi}_{\!\ssc C}^\dagger \rra_A$
is possible even with $A=0$. As mentioned above, the $A$ parameter has no role in
the original Lagrangian. Actually, Eq.(\ref{c2h}) also shows the parameter is
not constrained, while its value affects the determination of the auxiliary
component of ${\Phi}_{\!\ssc H}$.

We can see from above the ${\Phi}_{\!\ssc 0}$ and  ${\Phi}_{\!\ssc C}$ as superfield
composites have the quantum number of $\Phi_+\Phi_-$, while ${\Phi}_{\!\ssc H}$
has the conjugate quantum number. With the mass term generated, the two chiral 
superfields $\Phi_+$ and $\Phi_-$ make a Dirac pair. The simplest symmetry breaking 
picture of the SNJL model, for singlet superfields, is that of the Dirac fermion 
chiral symmetry, namely $U(1)_V \times U(1)_A \rightarrow U(1)_V$, which is the same 
as the original NJL model. The Standard Model with spontaneous electroweak symmetry 
breaking, however, has Weyl fermions as a basic ingredient with gauge symmetry 
forbidding any Dirac pairing. It is the same in  theories  with (N=1) supersymmetry.
Chiral superfields are chiral exactly because they bear each a
Weyl fermion. With such theories, chiral symmetry is not an issue. The
$U(1)_V \times U(1)_A$ symmetry is really a $U(1)_+ \times U(1)_-$,
or $\Phi_+$ and $\Phi_-$ number symmetries which one has no reason to
expect an interaction term to respect. It is broken by the dimension five
interaction in the HSNJL model. The SNJL model breaks the  $U(1)_+ \times U(1)_-$
symmetry dynamically to a $U(1)_+ - U(1)_-$ symmetry.

In the simplest model with the holomorphic four-superfield interaction discussed
above, presence of the ${\Phi}_{\!\ssc 0}^2$ term says that composite
${\Phi}_{\!\ssc 0}$ has to be in the real representation of the model symmetry.
Symmetry breaking options are hence very limited. In a model with more than two
basic chiral superfields (superfields $\Phi_+$ and/or $\Phi_-$ being multiplets), 
a holomorphic four-superfield interaction may give
rise to a wide range of symmetry breaking.  We focus in 
this paper in the simplest model of the type, to illustrate the basic feature.
To give an explicit symmetry breaking picture for the simplest version of the
HSNJL model (with singlet superfields), one can consider a $Z_4$ symmetry under 
which both the $\Phi_+$ and $\Phi_-$ superfields have a basic charge 
$e^{i\pi/2}$. The dimension five interaction obviously respects the symmetry
while the $\Phi_+\Phi_-$ Dirac mass term is not allowed, that is till the
 $Z_4$ symmetry is dynamically broken by the $\Phi_+\Phi_-$ vacuum condensate.
The condensate leaves a $Z_2$ symmetry that survives. An $SU(2)$ symmetry
breaking example can be constructed with $\Phi_+$ being an $SU(2)$-triplet
while having an equal and opposite $U(1)$ charge with a singlet $\Phi_-$. The  
$\Phi_+\Phi_-$ term is then invariant under the $U(1)$ but remains 
an $SU(2)$-triplet; its condensate breaks the $SU(2)$ symmetry. In that
case, the condensate will be along one of the three $SU(2)$ components. 
Hence the Dirac mass is only for the corresponding component of $\Phi_+$
which is what really forms a Dirac pair with the singlet $\Phi_-$. The other
two components of the $\Phi_+$ multiplet remains massless. Note that an 
explicit model giving rise to electroweak symmetry breaking in an effective
supersymmetric standard model has been discussed in Ref.\cite{034}.
The model involves three superfield multiplets with two superfield
condensates but otherwise the same holomorphic four-superfield interaction
structure. We will give more details for the continuous symmetry cases in
appendix B.

\section{Conclusion}
We presented above derivation of the superfield gap equations for the simplest
models with a dimension six or dimension five four-superfield interactions,
and analyzed some interesting cases for nontrivial solution. We proposed that
the gap equation in the superfield setting should be taken as one on the
Dirac mass parameter as a superspace scalar, like a constant superfield,
with both a supersymmetric and a supersymmetry breaking part. The amplitude
of the proper self-energy diagram, or two-point functions, effective action
$\dots$ etc., should all be considered in the same footing. Of course the
fermionic part should be zero. The supersymmetric breaking auxiliary part,
however, is in general nontrivial and could have an important role to play.
From our results for the Dirac mass parameter {$\mathcal{M}=m - \theta^2 \eta$}
in the case of the dimension five interaction, nontrivial solution for $m$
requires nonzero $\eta$, and vice versa. Considering the usual supersymmetric
Dirac mass $m$ only will completely miss the result. For the dimension six case,
though independent nontrivial solution for $m$ and $\eta$ are possible, a
nontrivial $\eta$ has its own interest and does affect the solution for $m$.
The approach will also be useful to check spontaneous supersymmetry breaking.

The two kinds of four-superfield interactions are alternative supersymmetrization
of the four-fermion interaction in the NJL model of dynamical mass generation
and symmetry breaking. They could each be used as a mechanism for dynamical
electroweak symmetry breaking. The two kinds of models (SNJL and HSNJL models)
have otherwise very different theoretical mass generation features, with
phenomenological implications. We illustrated some of the key aspects.

Our results explicitly establish the dynamical mass generation induced by
the dimension five four-superfield interaction, for the prototype HSNJL model.
The model has actually no four-fermion interaction and has a bi-scalar condensate,
instead of bi-fermion condensate, as the source of Dirac fermion mass. It has
otherwise theoretical features that look like a direct supersymmetric version
of the NJL model. It is arguably a more natural supersymmetrization of the
latter, though only constructed almost thirty years after the SNJL model
with the four-fermion interaction. It is expected to provide an alternative
paradigm for dynamical mass generation and symmetry breaking, at least for
superfield theories. The explicit symmetry breaking picture of the simplest
HSNJL model maybe considered as $Z_4 \rightarrow Z_2$. A version of the HSNJL 
with the basics superfields being (gauge) multiplets gives a simple application
to the breaking of a continuous symmetry. A simple example is given by an 
$SU(2)\times U(1)$ triplet and a singlet. We also have extended
versions with more than two basic superfield multiplets that 
can achieve a rich spectrum of dynamical
symmetry breaking. A case example, which was also the original target for
the idea of the HSNJL model is the one for electroweak symmetry
breaking. More details are available in appendix~B.

\acknowledgments
G.F. and O.K. are partially supported by research grant NSC 99-
2112-M-008-003-MY3, and G.F. further supported by grant NSC 99-2811-M-008-085
from the National Science Council of Taiwan.

\appendix
\section{Analysis of Conditions for Nontrivial Solutions of the Gap Equations}
We have two pairs of gap equations,
\bea
 m &=& 2 m  g^2  I_1(|m|^2, \tilde{m}^2, |\eta|, \Lambda^2) \;,
\nonumber \\
\eta &=& - \eta \, g^2 \tilde{m}_{\!\ssc C}^2 \, {I}_2(|m|^2, \tilde{m}^2, |\eta|, \Lambda^2) \;,
\label{Agap6}
\eea
for the case of a dimension six interaction, and
\bea
m &=&
\frac{\bar{\eta} G}{2}\; I_2(|m|^2,\tilde{m}^2,|\eta|, \Lambda^2)  \;,
\nonumber \\
\eta &=&  \bar{m} G \;  I_1(|m|^2,\tilde{m}^2,|\eta|, \Lambda^2)
- \frac{\bar{\eta} G B}{2} \; I_2(|m|^2,\tilde{m}^2,|\eta|, \Lambda^2) \;,
\label{Agap5}
\eea
for the case of the dimension five interaction.
\begin{figure}[!t]
\begin{center}
\begin{tabular}{cc}
{\includegraphics[height=6.0cm,width=6.0cm]{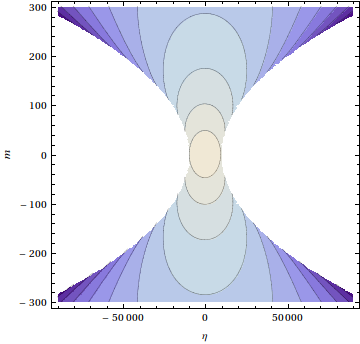}} &
\hspace*{.5in}
{\includegraphics[height=6.0cm,width=6.0cm]{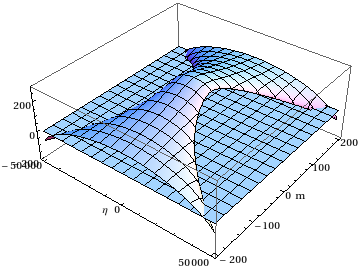}} \\
\end{tabular}
\end{center}
\caption{Contour plot (left) and 3D plot (right) of $I_1$ in the real
$\left(\eta,m\right)$ plane.}
\label{fig:I1}
\vspace*{.2in}
\hrule
\end{figure}
We first look into the general properties of the two loop-functions $I_1$ and $I_2$.
The shape of function $I_1$ is presented in Fig.~3 for generic given $\Lambda$ and
$\tilde{m}^2$. Here we consider only real values for $m$ and $\eta$. The blank regions
are where $|\eta| > m^2+ \tilde{m}^2$, which gives an unacceptable tachyonic mass for
a scalar mass eigenstate. We can see that the function is convex on the $\eta$-$m$
plane with its maximum value at the origin $\left(\eta=0,m=0\right)$. Along the
border lines that satisfy $|\eta|=m^2+\tilde{m}^2$, the value of function approaches
$-\frac{\Lambda^2}{32 \pi^2}$, definitely negative. The maximum value at the origin is
\bea
I_1\left(0,\tilde{m}^2,0,\Lambda^2\right)
= \frac{\tilde{m}^2}{16\pi^2}\log{\left[1+\frac{\Lambda^2}{\tilde{m}^2}\right]} \;.
\label{i1max}
\eea
On the $\eta$-axis ($m=0$), at the both ends, {\it i.e.}
$\left(|\eta|=\tilde{m}^2,0\right)$, we have
\bea
I_1\left(0,\tilde{m}^2,|\eta|=\tilde{m}^2,\Lambda^2\right) =
\frac{\tilde{m}^2}{16\pi^2}\log{\left[1+\frac{\Lambda^2}{2 \tilde{m}^2}\right]} \;.
\eea

The $I_2$ function, however, has the shape of a saddle as depicted in Fig.~4. It
is {concave} along the $\eta$-axis and {convex}  along the $m$-axis. It attains its
maximum value at $\left(|\eta|=\tilde{m}^2,0\right)$ on the centers of the
tachyonic exclusion borders ($|\eta| = m^2+ \tilde{m}^2$). The value is given be
\bea
I_2\left(0,\tilde{m}^2,|\eta|=\tilde{m}^2,\Lambda^2\right)=
\frac{1}{16\pi^2}\log{\left[1+\frac{\Lambda^2}{2 \tilde{m}^2}\right]} \;.
\label{i2max}
\eea
The origin is a local minimum, with a value given by
\bea
I_2\left(0,\tilde{m}^2,0,\Lambda^2\right)=
\frac{1}{16\pi^2}\left(\log{\left[1+\frac{\Lambda^2}{ \tilde{m}^2}\right]}
-\frac{\Lambda^2}{\Lambda^2+\tilde{m}^2}\right) \;.
\label{i2min}
\eea
In addition, it is positive definite.

With the gap equation set (\ref{Agap6}), nontrivial solution for $m$ requires
\bea
I_1=\frac{1}{2g^2} \;,
\eea
hence one can easily see the lower bound for $g^2$ from the maximum value of $I_1$ given
in Eq.(\ref{i1max}). The point corresponds to $\eta=0$. For any particular nonzero $\eta$,
an even larger $g^2$ will be required, but the simultaneous solution with
\bea
I_2=\frac{1}{-g^2 \tilde{m}_{\!\ssc C}^2} \;,
\eea
is required. The latter has a negative $\tilde{m}_{\!\ssc C}^2$ as a necessary condition.
The maximum value of $I_2$ within the no-tachyonic mass constraint given
in Eq.(\ref{i2max}) gives a lower bound for the magnitude of ${g^2 \tilde{m}_{\!\ssc C}^2}$
and at $m=0$ the local minimum given in Eq.(\ref{i2min}) gives an upper bound.
\begin{figure}[!t]
\begin{center}
\begin{tabular}{cc}
{\includegraphics[height=6.0cm,width=6.0cm]{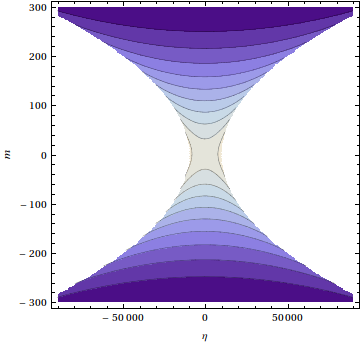}} &
\hspace*{.5in}
{\includegraphics[height=6.0cm,width=6.0cm]{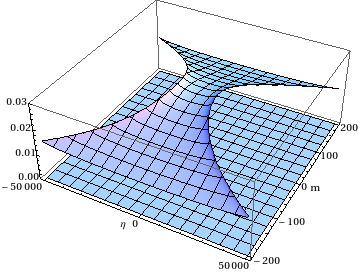}} \\
\end{tabular}
\end{center}
\caption{Contour plot (left) and 3D plot(right) of $I_2$ in the real $\left(\eta,m\right)$ plane.}
\label{fig:I2}
\vspace*{.2in}
\hrule
\end{figure}

With the gap equation set (\ref{Agap5}), one has to look for simultaneous nontrivial
solution for $m$ and $\eta$. Without lost of generosity, one can take $G$ to be real
and positive. For  $B=0$, the solution satisfies the equation
\bea
\frac{\eta^2}{2} I_2 = m^2 I_1 \;,
\eea
which is illustrated in Fig.~5 as (3) together with the two gap equations (1) and (2).
The general shapes of the three curves are independent of the values of $\Lambda$ and
$\tilde{m}^2$. We can see that so long as the slope of curve (1) at the origin is
larger than that of curve (2), existence of solution as given by the intersecting
points is to be expected. The slope of the curve (1) at the origin is given by
\bea
\frac{d m}{d \eta}_{\!\! (1)} = \frac{G}{32\pi^2}
    \left[\ln{\left(1+\frac{\Lambda^2}{\tilde{m}^2}\right)}
-\frac{\Lambda^2}{\tilde{m}^2+\Lambda^2}\right] \;,
\eea
and the slope of the curve (2) is given by
\bea
\frac{d m}{d \eta}_{\!\! (2)} =
\frac{1}{G}\frac{16\pi^2}{\tilde{m}^2\ln{\left(1+\frac{\Lambda^2}{\tilde{m}^2}\right)}} \;.
\eea
\begin{figure}[!t]
\begin{center}
{\includegraphics[height=7.0cm,width=8.0cm]{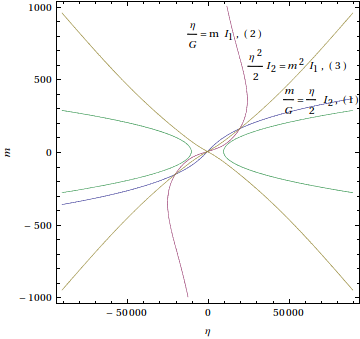}} \\
\end{center}
\caption{Solutions of to gap equations on the real $\left(\eta,m\right)$ plane.}
\label{fig:D5}
\vspace*{.2in}
\hrule
\end{figure}
Hence, the condition $\frac{d m}{d \eta}_{\!\!\ssc (1)}> \frac{d m}{d \eta}_{\!\!\ssc (2)}$ implies
\bea
G^2 > G_0^2
= \frac{512\pi^2}{\tilde{m}^2\ln{\left(1+\frac{\Lambda^2}{\tilde{m}^2}\right)}
              \left[\ln{\left(1+\frac{\Lambda^2}{\tilde{m}^2}\right)}
 -\frac{\Lambda^2}{\tilde{m}^2+\Lambda^2}\right]} \;.
\eea
For nonzero (real) value of $B$ small enough to be considered a perturbation for
the above case, the slope of the curve (2) is modified as
\bea
\frac{d m}{d \eta}_{\!\! (2)}
= \frac{1}{G}\frac{16\pi^2}{\tilde{m}^2\ln{\left(1+\frac{\Lambda^2}{\tilde{m}^2}\right)}}
  +\frac{B}{2}  \frac{\left[\ln{\left(1+\frac{\Lambda^2}{\tilde{m}^2}\right)}
-\frac{\Lambda^2}{\tilde{m}^2+\Lambda^2}\right]}
        {\tilde{m}^2\ln{\left(1+\frac{\Lambda^2}{\tilde{m}^2}\right)}} \;.
\eea
The condition for solution is modified to
\be
G > \sqrt{G_0^2 + b^2} + b \;,
\ee
where
\be
b = B \; \frac{8\pi^2}{\tilde{m}^2\ln{\left(1+\frac{\Lambda^2}{\tilde{m}^2}\right)}} \;\;.
\ee

\section{Application of the models to $SU(2)\times U(1)$ symmetry breaking}
We illustrate further in this appendix the application of the HSNJL
and the SNJL models to the phenomenological case of electroweak symmetry
breaking. The application is a key motivation for the construction of the
HSNJL model reported in Ref.\cite{034}. The reference discusses the effective
field theory picture, illustrates how the full Lagrangian for the MSSM can be
retrieved from an original model without the Higgs superfields. The only
interaction terms among the chiral superfields (besides the gauge
interactions) are the dimension five superpotential terms. What is
missing is the direct establishment of dynamical symmetry breaking
from a first principle gap equation analysis. Hence, we perform
the present study. Note that such a first principle establishment
of the symmetry breaking had not been available for the SNJL model
either.

To get the electroweak symmetry breaking of the MSSM, we need Higgs superfields
that are $SU(2)$ doublets. A direct application of the HSNJL model
only allows an effective Higgs in a real representation of the model
symmetry, hence not the doublet. However, getting two Higgs doublets
through a double composite/condensate structure is feasible \cite{034}. Before
going into that, let us illustrate first the single composite and single
condensate case of the SJNL model and a simpler $SU(2)\times U(1)$
symmetry breaking HSNJL model we have suggested in the main text.

For the case of the SNJL model, we have the basic Lagrangian
\bea {\cal L}  &=&
\int\!\!  d^4 \theta
  \left[ \hat{Q}^\dagger \hat{Q}(1-\tilde{m}^2_Q\theta^2 {\bar \theta}^2)
+\hat{T^c}^\dagger \hat{T^c}(1-\tilde{m}^2_t\theta^2 {\bar
\theta}^2)
\right]\nonumber \\
&+&   g^2 \int\!\! d^4 \theta \, \hat{Q}^\dagger \hat{T}^{c\dagger
} \hat{Q}\hat{T^c} \, (1+ \tilde{m}_C^2 \theta^2
\bar{\theta}^2)\,. 
\label{AdSL}\eea 
The superfield notation here
is the standard one in the MSSM, with $\hat{Q}$ being the quark
doublet superfield (containing $t_{\ssc L}$ and $b_{\ssc L}$) and
$\hat{T^c}$ the singlet one containing $\bar{t}$. For the
superfield gap equation analysis, we add the Dirac mass term
\bea
\int\!\!  d^4 \theta \left[ {\mathcal M}_t \, \hat{Q}\hat{T^c}
   \delta^2\!(\bar{\theta}) + h.c.     \right] \;,
\eea
and derive the superfield gap equation
for self-consistent solutions of ${\mathcal M}_t$ with
\bea
{\mathcal M}_t = m _t-{\theta}^2 \eta_t \;.
\eea
We need, in the derivation, the (hermitian conjugate of) following propagator
\bea
&& \lla T(\hat{Q}(1) \hat{T^c}(2)) \rra = \frac{i \,
\bar{m}_t}{p^2(p^2+|m_t|^2)} \frac{D_{\!\ssc 1}^2}{4}
\delta^4_{\ssc 12}
\nonumber \\&-&
\frac{i}{[(p^2+|m_t|^2+\frac{\tilde{m}_Q^2+\tilde{m}_t^2}{2})^2
-\frac{1}{4}(\tilde{m}_Q^2-\tilde{m}_t^2)^2-|\eta_t|^2]}
\! \left[ \frac{\bar{\eta}_t \, D_{\!\ssc 1}^2 \bar{\theta_{\!\ssc
1}}^2}{4} - \frac{\eta_t |m_t|^2 \, D_{\!\ssc 1}^2 {\theta_{\!\ssc
1}}^2}{4p^2} \!\right] \!\delta^4_{\ssc 12}
\nonumber  \\&+&
\frac{i\, \bar{m}_t \; [\tilde{m}_Q^2(p^2+|m_t|^2 +\tilde{m}_t^2)
-|\eta_t|^2]}{(p^2+|m_t|^2)[(p^2+|m_t|^2+\frac{\tilde{m}_Q^2+\tilde{m}_t^2}{2})^2-\frac{1}{4}(\tilde{m}_Q^2-\tilde{m}_t^2)^2-|\eta_t|^2]}
\; \frac{D_{\!\ssc 1}^2 \theta_{\!\ssc 1}^2 \bar{\theta_{\!\ssc
1}}^2}{4}
\nonumber  \\&+&
\frac{i\, \bar{m}_t \; [\tilde{m}_t^2(p^2+|m_t|^2
+\tilde{m}_Q^2)-|\eta_t|^2]}{(p^2+|m_t|^2)[(p^2+|m_t|^2+\frac{\tilde{m}_Q^2+\tilde{m}_t^2}{2})^2-\frac{1}{4}(\tilde{m}_Q^2-\tilde{m}_t^2)^2-|\eta_t|^2]}
\frac{\bar{\theta_{\!\ssc 1}}^2 \theta_{\!\ssc 1}^2 D_{\!\ssc
1}^2}{4} \! \delta^4_{\ssc 12} \;.
\eea
Note that the color indices are suppressed here, as in the Lagrangian.
\begin{figure}[!t]
\begin{center}
\includegraphics[height=4.0cm,width=10.0cm]{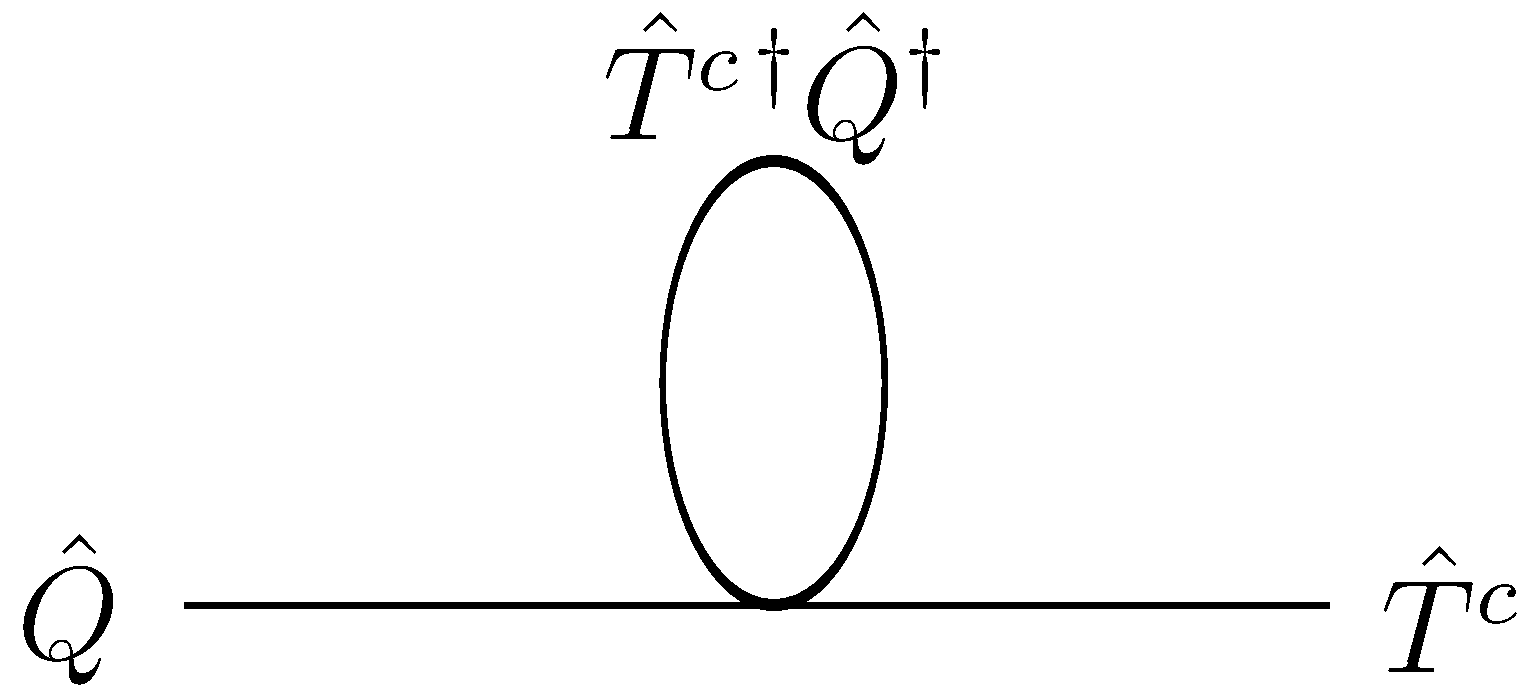}
\end{center}
\caption{\small Superfield diagram for proper self-energy
$\Sigma^t_{+-}(p, \theta^2)$ with the dimension six
four-superfield interaction.} \vspace*{.2in} \hrule
\label{figa1}\end{figure}
Following the calculations as illustrated in the main text
with $\Sigma^t_{+-}(p, \theta^2)$ shown in Fig.~\ref{figa1}, we have
\bea
m_t &=& 3\cdot 2~m_t~g^2~I'_1(|m_t|^2,\tilde{m}_Q^2,\tilde{m}_t^2,|\eta_t|, \Lambda^2) \;,
\nonumber \\
\eta_t &=& -3~\eta_t~g^2~\tilde{m}_C^2~
I'_2(|m_t|^2,
\tilde{m}_Q^2,\tilde{m}_t^2,|\eta_t|, \Lambda^2) \;,
\eea
where
\small
\bea
&& I'_1 \!\left(|m_t|^2, \tilde{m}_Q^2,\tilde{m}_t^2, |\eta_t|, \Lambda^2\right)
= \!\! \int\!\! \frac{d^4k}{(2\pi)^4} \frac{[\tilde{m}_Q^2(k^2+|m_t|^2+\tilde{m}_t^2)
+\tilde{m}_t^2(k^2+|m_t|^2+\tilde{m}_Q^2)-2|\eta_t|^2]}
{(k^2+|m_t|^2)[(k^2+|m_t|^2+\tilde{m}^2)^2-|\eta_t|^2]}
\nonumber \\ 
&=& \frac{1}{16\pi^2} \left[ \;\;  \frac{1}{2}
\left(|m_t|^2+\frac{\tilde{m}_Q^2+\tilde{m}_t^2}{2} \right) \,
\ln{\frac{(|m_t|^2+\tilde{m}_Q^2+\Lambda^2)(|m_t|^2+\tilde{m}_t^2+\Lambda^2)-|\eta_t|^2}{(|m_t|^2+\tilde{m}_Q^2)(|m_t|^2+\tilde{m}_t^2)-|\eta_t|^2}} \right.
 \nonumber \\  && 
- \;|m_t|^2 \ln{\frac{(|m_t|^2+\Lambda^2)}{|m_t|^2}} \;\;
+ \; \sqrt{|\eta_t|^2+\frac{1}{4}(\tilde{m}_Q^2-\tilde{m}_t^2)^2} \cdot
 \nonumber \\  && \left. 
\hspace{2cm}
\left(\tanh^{-1}\frac{|m_t|^2+\frac{\tilde{m}_Q^2+\tilde{m}_t^2}{2}+\Lambda^2}{\sqrt{|\eta_t|^2+\frac{1}{4}(\tilde{m}_Q^2-\tilde{m}_t^2)^2}}
-\tanh^{-1}\frac{|m_t|^2+\frac{\tilde{m}_Q^2+\tilde{m}_t^2}{2}}{\sqrt{|\eta_t|^2+\frac{1}{4}(\tilde{m}_Q^2-\tilde{m}_t^2)^2}}
\right) \;\;  \right] \;,
\nonumber \\
&&I'_2(|m_t|^2, \tilde{m}_Q^2,\tilde{m}_t^2, |\eta_t|, \Lambda^2) =
\int\frac{d^4k} {[(p^2+|m_t|^2+\frac{\tilde{m}_Q^2+\tilde{m}_t^2}{2})^2
-\frac{1}{4}(\tilde{m}_Q^2-\tilde{m}_t^2)^2-|\eta_t|^2]}
\nonumber \\
\hspace{-0.5cm}
&= &\frac{1}{16\pi^2} \left[ \frac{1}{2}
\ln{\frac{(|m_t|^2+\tilde{m}_Q^2+\Lambda^2)(|m_t|^2+\tilde{m}_t^2+\Lambda^2)-|\eta_t|^2}{(|m_t|^2+\tilde{m}_Q^2)(|m_t|^2+\tilde{m}_t^2)-|\eta_t|^2}}
\right. \nonumber \\
&\!\!\!\!\!+& \!\!\!\! \left.
\frac{|m_t|^2+\frac{\tilde{m}_Q^2+\tilde{m}_t^2}{2}}{\sqrt{|\eta_t|^2
+\frac{1}{4}(\tilde{m}_Q^2-\tilde{m}_t^2)^2}}
\left( \! \tanh^{-1} \! \frac{|m_t|^2+\frac{\tilde{m}_Q^2+\tilde{m}_t^2}{2}+\Lambda^2}
{{\sqrt{|\eta_t|^2+\frac{1}{4}(\tilde{m}_Q^2-\tilde{m}_t^2)^2}}}
-\tanh^{-1}\frac{|m_t|^2+\frac{\tilde{m}_Q^2+\tilde{m}_t^2}{2}}{{\sqrt{|\eta_t|^2+\frac{1}{4}(\tilde{m}_Q^2-\tilde{m}_t^2)^2}}}
\right) \!  \right] .
\nonumber\\
\eea
\normalsize
In the gap equation, there appears extra color factor of 3. The expression
corresponds otherwise exactly to the one given in the main text generalized to admit
unequal soft masses for the two superfields. After all, it is standard to apply
Feynman diagram calculations directly to nontrivial multiplets of any (gauge) symmetry.
More explicitly, as the $\hat{Q}\hat{T^c}$ combination in the Dirac mass term
${\mathcal M}_t \, \hat{Q}\hat{T^c}$ is an $SU(2)_L$ doublet, the ${\mathcal M}_t$
parameter is likewise a doublet vector. The $SU(2)_L$ symmetry is to be applied to
pick the nonzero direction of ${\mathcal M}_t$ as the symmetry breaking direction,
and the corresponding matching direction in $\hat{Q}\hat{T^c}$ may only then be
identified as the $t_{\ssc L}$ direction. With the direction assumed, the gap
equation is one for a superfield scalar parameter ${\mathcal M}_t$.

There is a simple, special, case for which it is easy to see that the gap equation
does admit nontrivial electroweak symmetry breaking solution. If we take
$\tilde{m}_Q^2=\tilde{m}_t^2$ in the Lagrangian of Eqn.(\ref{AdHL}), the case is
essentially the same as the one analyzed in the main text. Explicitly, the integrals
reduce exactly to the ones given there; and with the color factor 3 absorbed into
$g^2$ the gap equation becomes identical to the one analyzed. Hence, electroweak
symmetry breaking is established for the SNJL model. For the more general
$\tilde{m}_Q^2 \neq \tilde{m}_t^2$ case, solution analysis similar to the one for
the prototype case presented in the main text can be performed. One may also
consider modifying effects from a fully realistic Lagrangian, for example
from the QCD interaction. That is, however, beyond the scope of the present paper.

\begin{figure}[!t]
\begin{center}
\includegraphics[height=4.0cm,width=10.0cm]{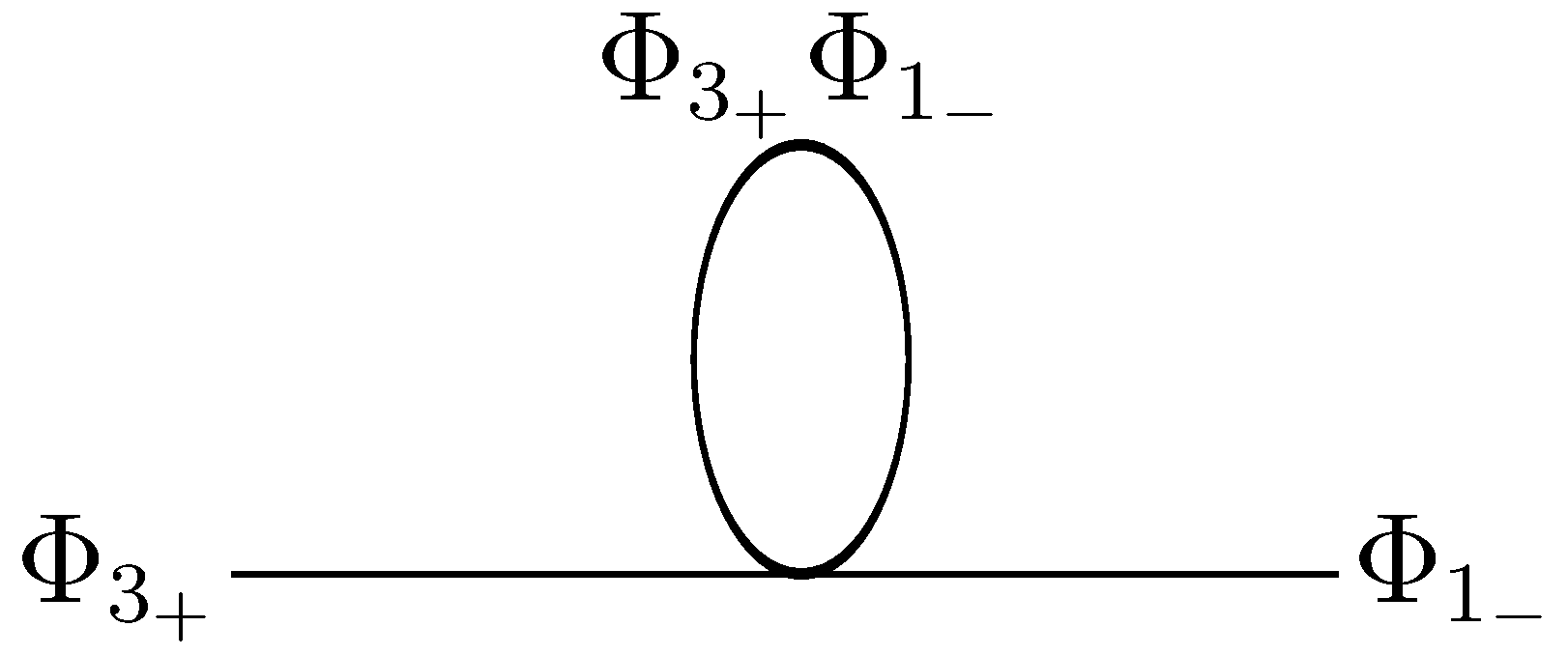}
\end{center}
\caption{\small Superfield diagram for proper self-energy
$\Sigma_{+-}(p, \theta^2)$ with the dimension five
four-superfield interaction involving a triplet and a singlet.} \vspace*{.2in} \hrule
\label{figa2}\end{figure}

Next, we take on the one composite/condensate HSNJL model with $SU(2)\times U(1)$
symmetry suggested in the main text.
This is a model with $\Phi_+ \equiv \Phi_{3_+}$, an $SU(2)$ triplet with
an $U(1)$ charge, and $\Phi_- \equiv \Phi_{1_-}$, an $SU(2)$ singlet
with the opposite $U(1)$ charge. We have, explicitly, the Lagrangian
\bea
{\cal L}  =
& &
\int\!\!  d^4 \theta ~
 \bigg[ \left( \Phi_{3_+}^\dagger \Phi_{3_+}
+ \Phi_{1_-}^\dagger \Phi_{1_-} \right) (1-\tilde{m}^2 \theta^2 {\bar \theta}^2)
 \bigg]
\nonumber \\
&-& \frac{G}{2} \int\!\!  d^4 \theta
\left( \Phi_{3_+} \Phi_{1_-} \right)^2 \,\delta^2\!(\bar{\theta})
+ h.c. \;,
\eea
in which we stick to the universal soft supersymmetry breaking mass term for
simplicity. By introducing the mass term ${\cal M} \Phi_{3_+} \Phi_{1_-}$,
with ${\cal M}=m-\theta^2 \eta$ as in the main text, the supergraph calculation
and resulted gap equation as derived from the diagram in Fig.~\ref{figa2} are
exactly those given in the main text. Again, the Dirac mass ${\cal M}$ is naively
an $SU(2)$ triplet vector in which one direction would be single out by the
symmetry breaking. Hence, our analysis in the main text leads to the conclusion
that the HSNJL model has $SU(2)\times U(1)$ symmetry breaking solutions.
The model can be extended to have the symmetry as a gauge one. It may be
consider a prototype model of continuous (gauge) symmetry breaking with the
dimension five four-superfield interaction. However, the composite superfield
$\Phi_{3_+} \Phi_{1_-}$ developing vacuum condensate is an $SU(2)$ triplet
with zero charge under the $U(1)$, hence not one that can be used for the
phenomenological electroweak symmetry breaking.

The HSNJL case to be applied to the MSSM is a bit more complicated.
The basic Lagrangian is
\bea {\cal L}  &=&
\int\!\!  d^4 \theta ~
  \left[ \hat{Q}^\dagger \hat{Q}(1-\tilde{m}^2_Q\theta^2 {\bar \theta}^2)
+\hat{T^c}^\dagger \hat{T^c}(1-\tilde{m}^2_t\theta^2 {\bar
\theta}^2) +\hat{B}^{c\!^\dagger} \hat{B^c}(1-\tilde{m}^2_b\theta^2
{\bar \theta}^2) \right]\nonumber \\
&-&\frac{G}{2} \int\!\! d^4 \theta \,
\hat{Q}\hat{T^c}\hat{Q}\hat{B^c} \, (1+ B \theta^2)\,
\delta^2\!(\bar{\theta}) + h.c. \;.
\label{AdHL}\eea
\begin{figure}[!t]
\begin{center}
\subfigure[]{\includegraphics[height=4.0cm,width=6.0cm]{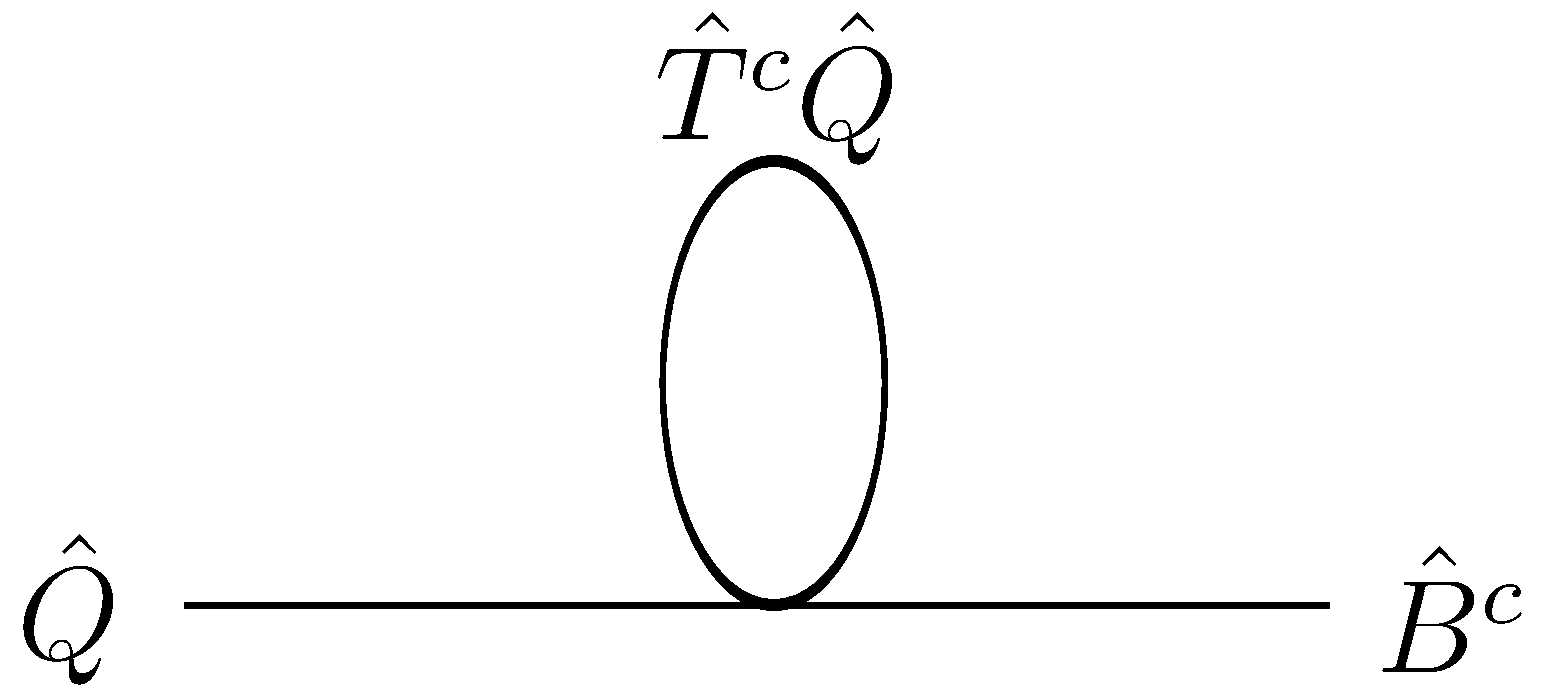}}
\hfill
\subfigure[]{\includegraphics[height=4.0cm,width=6.0cm]{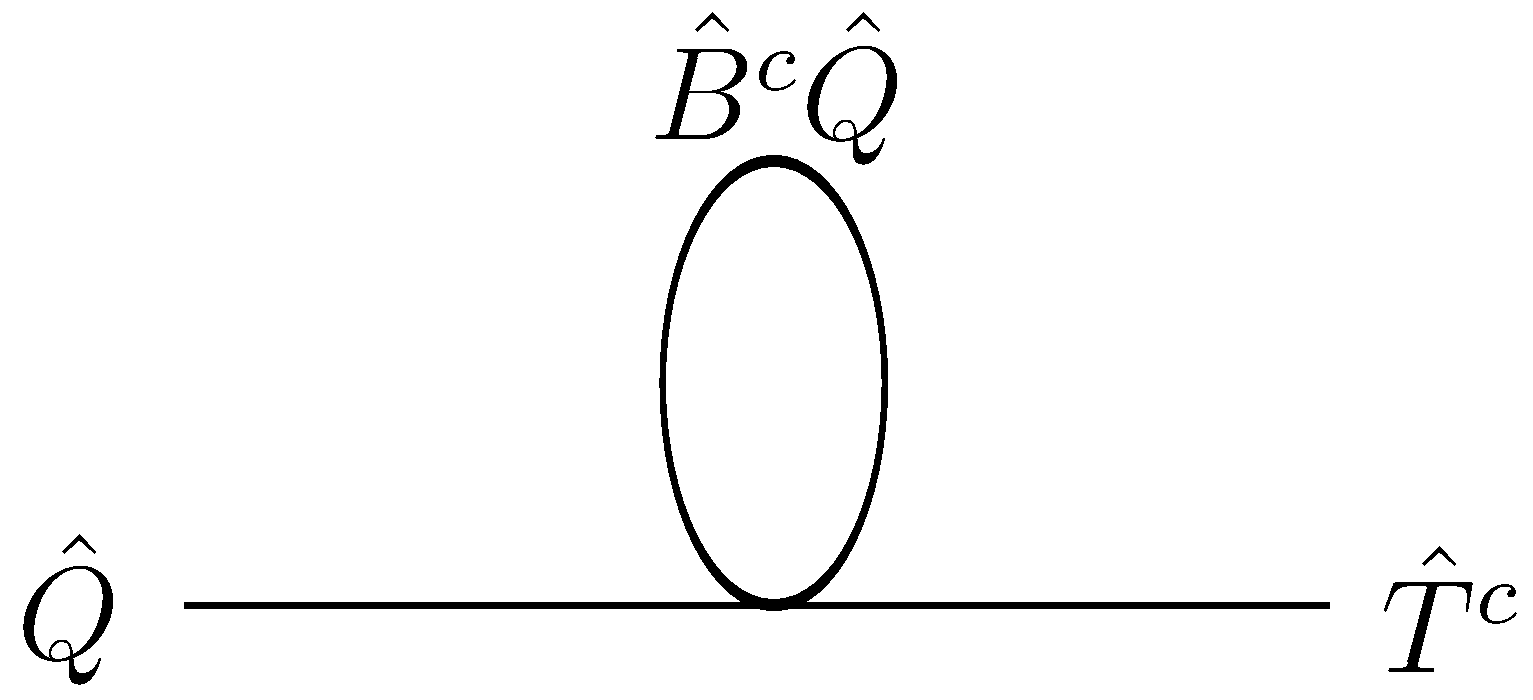}}
\end{center}
\caption{\small Superfield diagrams for proper self-energy,
 (a) for $\Sigma^b_{+-}(p, \theta^2)$ and
(b) for $\Sigma^t_{+-}(p, \theta^2)$, with the dimension five
four-superfield interaction.} \vspace*{.2in} \hrule
\label{figa3}\end{figure}
The basic Lagrangian, in the presence of $\hat{Q}\hat{T^c}$ and $\hat{Q}\hat{B^c}$
composites as the effective $\hat{H}_d$ and $\hat{H}_u$, respectively, can be
easily extended to give the full MSSM Lagrangian as the effective theory \cite{034}.
For the superfield gap equation analysis, we have to consider the two Dirac mass terms
\bea
\int\!\!  d^4 \theta ~ \left[({\mathcal M}_t \, \hat{Q}\hat{T^c}+{\mathcal M}_b \,
\hat{Q}\hat{B^c}) \,\delta^2\!(\bar{\theta}) + h.c. \right] \;,
\eea
with
\bea
&&{\mathcal M}_t = m _t-{\theta}^2 \eta_t \;,
\nonumber \\
&&{\mathcal M}_b = m _b-{\theta}^2 \eta_b \;.
\eea
We need the propagator $\lla T \left(\hat{Q}(1) \hat{T^c}(2)\right) \rra$
given explicitly above together with an exact matching one for
$\lla T \left(\hat{Q}(1) \hat{B^c}(2)\right) \rra$ (with all subscript $t$
replaced by $b$). Following the calculations as illustrated in the main text
with $\Sigma^b_{+-}(p, \theta^2)$ shown in Fig.~\ref{figa3}, we have
\bea
m_b &=& 2 \frac{\bar{\eta}_t G}{2}\; I'_2(|m_t|^2,
\tilde{m}_Q^2,\tilde{m}_t^2,|\eta_t|, \Lambda^2)\;,
\nonumber \\
\eta_b &=&  2 \bar{m}_t G \;
I'_1(|m_t|^2,\tilde{m}_Q^2,\tilde{m}_t^2,|\eta_t|, \Lambda^2) -
2 \frac{\bar{\eta}_t G B}{2} \; I'_2(|m_t|^2,
\tilde{m}_Q^2,\tilde{m}_t^2,|\eta_t|, \Lambda^2) \;.
\label{gb}\eea
Similarly, $\Sigma^t_{+-}(p, \theta^2)$ shown in Fig.~\ref{figa3} gives
\bea
m_t &=& 2 \frac{\bar{\eta}_b G}{2}\; I'_2(|m_b|^2,
\tilde{m}_Q^2,\tilde{m}_b^2,|\eta_b|, \Lambda^2)\;,
\nonumber \\
\eta_t &=&  2 \bar{m}_b G \;
I'_1(|m_b|^2,\tilde{m}_Q^2,\tilde{m}_b^2,|\eta_b|, \Lambda^2) -
2 \frac{\bar{\eta}_b G B}{2} \; I'_2(|m_b|^2,
\tilde{m}_Q^2,\tilde{m}_b^2,|\eta_b|, \Lambda^2) \;,
\label{gt}\eea
[where $I'_1(|m_t|^2,\tilde{m}_Q^2,\tilde{m}_t^2,|\eta_t|,
\Lambda^2)$ and $I'_2(|m_t|^2,\tilde{m}_Q^2,\tilde{m}_t^2,|\eta_t|,
\Lambda^2)$ can be obtained from
$I'_1(|m_b|^2,\tilde{m}_Q^2,\tilde{m}_b^2,|\eta_b|, \Lambda^2)$ and
$I'_2(|m_b|^2,\tilde{m}_Q^2,\tilde{m}_b^2,|\eta_b|, \Lambda^2)$ by
the replacements  $m_b$ by  $m_t$, $\tilde{m}_b^2$ by $\tilde{m}_t^2$
and $\eta_b$ by $\eta_t$]. In both cases there is extra factor 2
which appears from the color factor. (In fact, the interaction term
with the color indices reads
$\hat{Q}^\za \hat{T}^c_\za \hat{Q}^\zb \hat{B}^c_\zb$
without which the indistinguishable $\hat{Q}^\za$ and $\hat{Q}^\zb$
giving vanishing result as the singlet direction is
antisymmetric in the $SU(2)$ indices of the two $\hat{Q}$ doublets
--- hence the factor 2 instead of 3.)

We have the model gap equations as given by Eqns.(\ref{gt}) and
(\ref{gb}). Nontrivial solutions to the two superfield Dirac mass
parameters ${\mathcal M}_t$ and ${\mathcal M}_b$ correspond to
electroweak symmetry breaking with two $SU(2)\times U(1)$ doublets
aligned to preserve the electromagnetic $U(1)$. We have to leave
reporting nontrivial solutions for the generic case to a future
publication, due to the very demanding analysis involved. However,
we can again use a simple, special, case to establish that
the usual electroweak symmetry breaking can be obtained. If we
take $\tilde{m}_b^2=\tilde{m}_t^2$ in the original
Lagrangian of Eqn.(\ref{AdHL}), the model dynamic is obviously
symmetrical for $t$ and $b$. That naturally suggests looking for
solution with ${\mathcal M}_t={\mathcal M}_b$. In this case, the
two set of gap equations collapsed into one. Take further the
same soft mass value for $\tilde{m}_Q^2$. The set of gap equations
then becomes identical to the one of our prototype model discussed
in the main text, with nontrivial solution explicitly illustrated.
Under the special case, we have electroweak symmetry breaking for
the MSSM, with however phenomenologically wrong identical top and
bottom masses. To get the right masses, we sure need
 $\tilde{m}_b^2 \ne \tilde{m}_t^2$ as a starting point.

Our key purpose here is to illustrate explicitly the HSNJL model as
one that is capable of giving rising to dynamical symmetry breaking
and Dirac mass generation, including interesting continuous
symmetry like $SU(2)\times U(1)$. The most phenomenologically
interesting application would be for the case with the MSSM
as the effective field theory, which we also discussed here.
The model can also be easily applied to other symmetry breaking
setting of possible phenomenological interest,
such as a grand unification symmetry.

\def\PRD #1 #2 #3 {Phys. Rev. D {\bf#1},\ #2 (#3)}
\def\PRL #1 #2 #3 {Phys. Rev. Lett. {\bf#1},\ #2 (#3)}
\def\PLB #1 #2 #3 {Phys. Lett. B {\bf#1},\ #2 (#3)}
\def\NPB #1 #2 #3 {Nucl. Phys. {\bf B#1},\ #2 (#3)}
\def\ZPC #1 #2 #3 {Z. Phys. C {\bf#1},\ #2 (#3)}
\def\EPJ #1 #2 #3 {Euro. Phys. J. C {\bf#1},\ #2 (#3)}
\def\JHEP #1 #2 #3 {JHEP {\bf#1},\ #2 (#3)}
\def\IJMP #1 #2 #3 {Int. J. Mod. Phys. A {\bf#1},\ #2 (#3)}
\def\MPL #1 #2 #3 {Mod. Phys. Lett. A {\bf#1},\ #2 (#3)}
\def\PTP #1 #2 #3 {Prog. Theor. Phys. {\bf#1},\ #2 (#3)}
\def\PR #1 #2 #3 {Phys. Rep. {\bf#1},\ #2 (#3)}
\def\RMP #1 #2 #3 {Rev. Mod. Phys. {\bf#1},\ #2 (#3)}
\def\PRold #1 #2 #3 {Phys. Rev. {\bf#1},\ #2 (#3)}
\def\IBID #1 #2 #3 {{\it ibid.} {\bf#1},\ #2 (#3)}

\end{document}